\RequirePackage{etex}
\documentclass{aa}
\usepackage{textcomp}
\usepackage{graphicx}
\usepackage{txfonts}
\usepackage{xcolor}
\usepackage{gensymb}
\begin{document}

   \title{EWOCS-VI: Probing the hidden intermediate-mass population of Westerlund 1\thanks{Corresponding author: \email{ccordenes@uc.cl}}}
   \titlerunning{EWOCS-VI}

   \author{C. Ordenes-Huanca
          \inst{1,2}, A. Bayo\inst{3}, M. Guarcello\inst{4}, M. Zoccali\inst{1},
          A. Rojas-Arriagada\inst{5,6},
          V. Almendros-Abad\inst{4},      M.~Andersen\inst{3},
          R. Bonito\inst{4},
          V. Guzmán\inst{1,7},
          E. Moraux\inst{8},
          K. Muzic\inst{9},
          L. Prisinzano\inst{4},
          S. Sciortino\inst{4},
          \and N.\,J.~Wright\inst{10}
          }
    \authorrunning{Ordenes-Huanca et al.}

   \institute{Instituto de Astrofísica, Pontificia Universidad Católica de Chile, Casilla 306, Santiago 7820436, Chile
        \and
            Departamento de Astronomía, Universidad de Concepción, Casilla 160-C, Concepción, Chile
        \and
             European Southern Observatory, Karl-Schwarzschild-Strasse 2, 85748 Garching bei München, Germany
        \and
            Istituto Nazionale di Astrofisica (INAF) – Osservatorio Astronomico di Palermo, Piazza del Parlamento 1, 90134 Palermo, Italy
        \and
            Departamento de Física, Universidad de Santiago de Chile, Av. Víctor Jara 3659, 9170124, Santiago, Chile
        \and
            Center for Interdisciplinary Research in Astrophysics and Space Exploration (CIRAS), Universidad de Santiago de Chile, Santiago, Chile
        \and Millennium Nucleus on      Young Exoplanets and         their Moons (YEMS),    Chile
        \and
            Université Joseph-Fourier Grenoble 1 / CNRS-INSU, Institut de Planétologie et d’Astrophysique de Grenoble (IPAG) UMR 5274, 38041 Grenoble, France
        \and
            Instituto de Astrofísica e Ciências do Espaço, Faculdade de Ciências, Universidade de Lisboa, Ed. C8, Campo Grande, 1749-016 Lisbon, Portugal
        \and 
            Astrophysics Group, Keele University, Keele ST5 5BG, United Kingdom    
             }

   \date{Received 24 April 2026 / Accepted 24 June 2026}

  \abstract
   {Westerlund 1 (Wd1), the most massive young star cluster in the Milky Way, is an excellent laboratory for studying star formation and early stellar evolution in a starburst-like environment. However, high extinction restricts studies of its stellar content, and focus on high-mass stars limits our knowledge of the full spatial extent of the cluster.}
   {We characterize the near-infrared (NIR) variability of the stellar population of Wd1, filling the mass gap between massive stars traced by \textit{Gaia} and very low-mass stars from previous Extended Westerlund 1 and 2 Open Clusters Survey (EWOCS) studies, to provide a more complete view of cluster membership across solar and super-solar masses.}
   {We exploited data from the VISTA Variables of the Vía Láctea survey and its extension (VVVX), using NIR point spread function (PSF) photometry and astrometric solutions from its latest data release, namely the VIRAC2 catalogs, mainly in the $K_{\rm s}$ band. Their large spatial coverage enables study of both the central regions and outskirts of the cluster. We applied {\tt HDBSCAN} clustering algorithm in a 6D parameter space to differentiate cluster members from field contaminants, assessing robustness through Monte Carlo simulations. Variable sources along the line of sight were also identified and characterized.}
   {We identify $1286$ high-probability candidate members ($12 \leq J \leq 18\,\rm{mag}$) spanning $\sim 1.5$--$20\,M_{\odot}$, adopting both PARSEC $5$ and $6\,\rm{Myr}$ isochrones ($A_{K_{\rm{s}}}=0.6\,\rm{mag}$, $d=4.23\,\rm{kpc}$). A considerable fraction ($\sim 34\%$) shows statistically significant flux variations. We present, for the first time, a parametric analysis of variability modes of Wd1 candidate members in the $K_{\rm{s}}$ band, providing a membership catalog suitable for future kinematic studies.}
   {}


   \maketitle
%

\section{Introduction}

The near-infrared (NIR) view of our own Galaxy allows us to study its heavily extincted regions. It has become one of the main strategies to limit the effects of dust and to observe and analyze otherwise hidden stellar populations. In the last decades, the VISTA Variables of the Via Láctea (VVV Survey) and its extension VVVX (hereafter VVVX, \citeauthor{minniti+2010} \citeyear{minniti+2010}; \citeauthor{Saito_2024} \citeyear{Saito_2024}) allowed us to observe stars located near the Galactic plane, in the Bulge, and the southern disk of the Milky Way. Its photometric measurements, in $Y$, $Z$, $J$, $H$, and $K_{\rm{s}}$ bands, from 2009 to early 2024, have given us access to several of the most extincted areas in the Galaxy, such as star-forming regions, which can contain thousands of young stars hidden by the typically uneven presence of dust. Also, it has enabled us to explore, in some cases for the first time, the variable population of these regions through $K_{\rm{s}}$ band observations \citep{OrdenesHuanca_2022, Ordenes-Huanca_2024}. Its latest data release, the VVV Infrared Astrometric Catalogue version 2 (VIRAC2) contains point-spread-function photometry with a completeness at a level of $90\%$ for sources in the magnitude range of $11\leq K_{\rm{s}}\leq 16\,\rm{mag}$. It can reach up to $K_{s}\approx 17.5\,\rm{mag}$ in most fields \citep{Smith_2025} and also provides time-series data from a $5$-year baseline (from 2010 to 2015), parallaxes and proper motions (PMs) for more than $545$ millions of stars.\\

Regarding high extinction and extreme star formation history, Westerlund 1 (Wd1), represents a very interesting and challenging case to exploit the VVVX data. Wd1 is the most massive young star cluster known in the Milky Way \citep{Clark_2005} with mass estimates ranging between $10^{4}$ and $10^{5}\,M_{\odot}$. It is centrally dominated by numerous high-mass stars, which include OB stars \citep{Negueruela_2010} and Wolf-Rayet sources, yellow and red supergiants, and luminous blue variables \citep{Clark_2005}. Beyond this crowded core region of radius of about $3.5\,\rm{arcmin}$, a halo of possible cluster members has been identified \citep{Negueruela_2022, Guarcello_2024}. Thus, two populations, one confined near the cluster center and the other extended across a larger area, coexist in Wd1. They contain stars of various masses where the more massive and central ones influence the evolution of those less massive and more spread \citep{Guarcello_2025}.\\

\textit{Gaia} astrometry \citep{Gaia_2018, Gaia_2023} has provided important constraints on the distance and bulk PM of Wd1 using its most massive members, mainly in the range $\geq 30\,M_{\odot}$ \citep{Clark_2010, Negueruela_2022}. The \textit{Gaia}-based census by \citet{Negueruela_2022} reaches down to $\sim 15\,M_{\odot}$ and can complement the VLT/FLAMES ($R\sim16\,200$) spectroscopic sample by \citet{Clark_2020}, which  extends to at least $\sim 20\,M_{\odot}$. Still, both are significantly incomplete at these limits. The strong extinction toward the cluster (between $\sim 9$ and $15$\,mag according to \citet{Damineli_2016}), and the distance at which Wd1 is located ($\sim4$\,kpc, according to \citet{Negueruela_2022}, \citet{Lim_2013}, and \citet{Beasor_2021}) severely limit the completeness of \textit{Gaia}-based samples, motivating the use of NIR data to extend membership studies to fainter and intermediate-mass stars of $M<15\,M_{\odot}$.\\

The Extended Westerlund 1 and 2 Open Clusters Survey (EWOCS\footnote{\url{https://westerlund1survey.wordpress.com/}}) is significantly advancing our understanding of these clusters by extending our census of their members to much lower masses, up to $0.06\,M_{\odot}$. Its main goal is to study  the formation and evolution of stars and their possible planets located in a starburst environment \citep{Guarcello_2024}. The survey focuses on the analysis of the two closest supermassive star clusters, Wd1 and Westerlund 2 (Wd2). For the former, X-ray observations, using \textit{Chandra}, combined with IR data have probed its low to intermediate mass content from $0.8$ to $2\,M_{\odot}$, while James Webb Space Telescope (JWST) observations have reached its brown dwarf regime down to $0.06\,M_{\odot}$ \citep{Guarcello_2024, Guarcello_2025}.\\

NIR studies have included the analysis of stars of $\sim 25\,M_{\odot}$ \citep{Gennaro_2011}, using NTT/SofI data, and even down to $0.15\,M_{\odot}$ \citep{Andersen_2017} from HST/WFC3. On the other hand, since X-rays require an optical or IR counterpart to resolve the stellar photosphere and determine stellar properties, and the optical wavelength range is strongly limited by extinction, a gap remains between the two mass regimes probed by each EWOCS dataset. This gap has prevented us from analyzing the Wd1 members in the most comprehensive way, which is fundamental to avoid possible bias in the obtained parameters—these parameters could be strongly limited by the mass range used. Therefore, IR studies are crucial for identifying members in this mass gap and to use the entire mass range to verify the cluster's physical parameters.\\

Whereas the JWST observations saturate on intermediate mass sources ($\gtrsim 3\,M_{\odot}$) even in the shortest exposures \citep{Guarcello_2025}, VVVX, with its shallower effective depth, does not suffer from saturation in this mass range, making the two datasets naturally complementary. Conversely, in the innermost region of the cluster (within about $2\,\rm{arcmin}$), the massive and luminous stars that dominate the core saturate the VVVX images, leaving very few usable photometric detections there. Together, these limitations define the mass and spatial window that VVVX can uniquely probe: intermediate-mass stars outside the crowded cluster core.\\

The main goal of this study is to bridge the census on the aforementioned mass gap through an astrometry-based analysis. The astrometric census presented here provides the basis for a detailed analysis of the internal kinematics and dynamical state of Wd1, which is explored in a companion paper (Ordenes-Huanca et al. in prep.) and used to constrain its formation pathway. This study complements previous and recent works that have probed the region in complementary mass regimes \citep{Negueruela_2022, Guarcello_2024, Guarcello_2025}. Moreover, by using VVVX and its most recent data release, the VIRAC2 catalog, we can examine a larger region for identifying members. This will let us explore the cluster halo, its environment, and its dispersed population, which are key to understanding cluster formation and future evolution.\\

Variability studies of Wd1 members have also been conducted, but these have focused only on the most massive stars (\citeauthor{Bonanos_2007} \citeyear{Bonanos_2007}; \citeauthor{Clark_2010} \citeyear{Clark_2010}). In this study, we examine the variability of $\sim 1.5 $--$20\,M_\odot$ cluster members for the first time. Variability of members in this mass range is very challenging to study due to the faintness of the targets and their associated photometric errors. The latter are expected to increase for fainter objects, in which case, the variability will be hard to detect. Variability detection is also limited to the sampling and cadence of the data, the amplitude of the flux changes, and the timescale of the observations. VVVX is very suitable for the latter due to its $12$ years baseline. Indeed, it has proven to be an important tool for finding pre-main sequence periodic bursts \citep{Guo_2022} and spot modulated changes \citep{OrdenesHuanca_2022} that are apparent on shorter timescales.\\

This paper is structured as follows. Sect.~\ref{selection} is devoted to the description of the method considered to identify candidate members of Wd1. The photometric features of the proposed census are analyzed in Sect.~\ref{candidates}, whereas Sect.~\ref{variability} focuses on variability. We compare our findings with the literature and discuss their implications in Sect.~\ref{discussion} and summarize our conclusions in Sect.~\ref{conclusions}.\\

\section{Analysis and results}
\subsection{Member selection from VIRAC2 data}
\label{selection}

The NIR data considered in this work come from the VVVX Survey \citep{minniti+2010}, which monitored the sky between 2009 and 2024, giving a unique timescale of observation, particularly in the $K_{\rm{s}}$ band, which comprises the main part of its time series data. Using an uneven observation sampling, VIRAC2 time series catalog provides a higher cadence from 2010 to 2015. The frequency of observations ranges from tens per year to daily observations and allows to find flux changes from a fraction of a day to years.\\

\subsubsection{Search radius and {\tt HDBSCAN} clustering parameters}

To start an initially blind search of members, we retrieve all VIRAC2 data around the Wd1 center, $\alpha$ (J2000) $=251.76\,\rm{deg}$ and $\delta$ (J2000) $=-45.85\,\rm{deg}$ from the ESO Catalogue facility\footnote{\url{https://www.eso.org/qi/catalogQuery/index/424 }}. The search radius was sampled to balance completeness against contamination. The search radius used to retrieve the VIRAC2 data was optimized according to previous determinations of the cluster’s extent. For instance, \citet{Dias_2002} measured a cluster radius of $1.2\,\rm{arcmin}$, while \citet{Negueruela_2022} pointed to an elliptical shape of the cluster with a major axis of $7\,\rm{arcmin}$. Radius measurements can be controversial because they depend on the definition, the method to compute them, and the bandpass of observation \citep{Sanchez_2020}. \\

Considering the aforementioned values, the membership search was conducted with a range of radii $R_{s}$ from $5\,\rm{arcmin}$ to $17\,\rm{arcmin}$, in steps of $1\,\rm{arcmin}$. Due to VVVX saturation limits, the very center of the cluster, within a radius of about $2\,\rm{arcmin}$, has very few usable photometric detections, since the most massive members are saturated and therefore are not included in the VIRAC2 source catalog \citep{Smith_2025}. To identify possible new intermediate to low-mass members of Wd1, we considered photometric and kinematic data of VIRAC2 of the sources within each search radius. We applied a {\tt HDBSCAN} clustering algorithm (\citeauthor{Campello2013}, \citeyear{Campello2013}; \citeauthor{McInnes2017}, \citeyear{McInnes2017}) to the normalized data set.  In particular, using the equatorial coordinates of the sources $\alpha$ and $\delta$, their PM values $\mu_{\alpha}$, $\mu_{\delta}$ and the $J-K_{\rm{s}}$ color of each.\\

We did not include the parallax measurements from the VIRAC2 data. The main reason is due to their large uncertainties, which can be one order of magnitude higher than the expected parallax of the Wd1 cluster members. For example, \citet{Negueruela_2022} found a mean parallax for the cluster members of $\varpi=0.238 \pm 0.012\,\rm{mas}$, while the mean VIRAC2 parallax errors in the region surrounding Wd1 are $\sim 10\,\rm{mas}$. This limitation applies specifically to the parallaxes and should not be interpreted as a limitation of the full astrometric solution. In the magnitude range shared with \textit{Gaia}, the VIRAC2 PMs are consistent, within uncertainties, with previous \textit{Gaia}-based determinations for Wd1 (see also Sect.~3.2). For this reason, we excluded parallaxes from the clustering features, but retained PMs as a key ingredient of the membership selection.\\

There are also discrepancies in the age determinations of Wd1. \citet{Beasor_2021} proposed a continuous star formation process in the cluster. Pre-main sequence stars have been pointed to be in the age range of $7$--$10\,\rm{Myr}$, and a similar value was obtained by \citet{Navarete_2022}. Nevertheless, several studies seem to be in agreement with a younger age for Wd1 of about $5\,\rm{Myr}$ (e.g. \citeauthor{Clark_2005} \citeyear{Clark_2005}, \citeauthor{Negueruela_2010} \citeyear{Negueruela_2010}, \citeauthor{Kudryavtseva_2012} \citeyear{Kudryavtseva_2012}). For consistency with those works, we include a $5\,\rm{Myr}$ isochrone as a reference case. However, more recent analyses suggest a somewhat older age, closer to $6\,\rm{Myr}$ \citep{Castellanos_2026}, and we therefore also consider this age value in our study. The same $R_{s}$ analysis for a PARSEC isochrone of $6\,\rm{Myr}$ was developed separately. Both isochrones were placed at a distance of $4.23\,\rm{kpc}$ \citep{Negueruela_2022}, using an extinction value $A_{K_{\rm{s}}}=0.69\,\rm{mag}$ compatible with the one of \citet{Negueruela_2010}.\\

Furthermore, we defined a sixth parameter called \textit{min\_dist}, which indicates the distance to each point in the $J$ vs. $J-K_{\rm{s}}$ color-magnitude diagram (CMD) from a PARSEC isochrone, including only main sequence and pre-Main Sequence stages. Once an isochrone was placed in the CMD, we computed the \textit{min\_dist} parameter for all stars with available $J$ and $K_{\rm{s}}$ photometry in VIRAC2 data. This metric was also included in the {\tt HDBSCAN} clustering algorithm to find possible clusters in the region. The latter was run separately adopting $5\,\rm{Myr}$ and $6\,\rm{Myr}$ isochrones. Therefore, the six parameters involved in the {\tt HDBSCAN} runs are: $\alpha$ and $\delta$ equatorial coordinates, PM values $\mu_\alpha \cos{\delta}$, $\mu_{\delta}$, $J-K_{\rm{s}}$ color, and the distance to the isochrones considered in the $J$ vs. $J-K_{\rm{s}}$ CMD.\\

To choose a proper search radius, we investigated a range of $R_s$ values. We applied these parameters and ran the {\tt HDBSCAN} clustering algorithm in all the range of search radii. The \textit{min\_cluster\_size} was set to $150$ and the normalization was carried out using a robust scaler \citep{Pedregosa_2011}. Each of the runs always returned two clusters: one with PMs expected for Wd1 members (approximately centered on $\mu_\alpha \cos{\delta}=-2.3\,\rm{mas\,yr^{-1}}$ and $\mu_{\delta}=-2.7\,\rm{mas\,yr^{-1}}$) and other with PMs more associated with the field (around $\mu_\alpha \cos{\delta}=-3.1\,\rm{mas\,yr^{-1}}$ and $\mu_{\delta}=-4.3\,\rm{mas\,yr^{-1}}$).\\

\subsubsection{Clustering robustness against uncertainty in the data}

As mentioned before, the features fed to the clustering algorithm did not include uncertainties. The expectation is that these uncertainties are highly correlated, and among them, the PMs exhibit the largest relative values, ranging from $0.05$ to $8$. To asses the impact of this heterogeneity in our results, we performed a simple Monte Carlo simulation: we drew $50$ populations compatible with our base one assuming Gaussian distributions for the PMs (we did not sample any other feature). Each population was analyzed in the same way described in the previous subsection, yielding a distribution of probabilities for cluster memberships per object.

\subsubsection{Membership criteria}

The metrics derived from the Monte Carlo experiment, probabilities (\textit{prob}) and errors (\textit{std}), were used to define a relative error $rel\equiv std/prob$. Several thresholds to \textit{rel} were defined to distinguish degrees of certainty for candidates. The first defined class was named \textit{core stars}, which contains stars with $prob - std \geq 0.8$ and $rel \leq 0.3$. These sources are the most robust cluster member candidates in our sample and maintain high probability even when accounting for probability uncertainties. On the other hand, a second class was denominated \textit{high-confidence stars}, which was defined to have sources showing $prob - std \geq 0.6$, 
$prob \geq 0.8$ and $rel \leq 0.4$. The latter have a high membership probability but slightly higher uncertainty than core candidate members.\\

\begin{figure}
    \centering
    \includegraphics[width=0.47\textwidth]{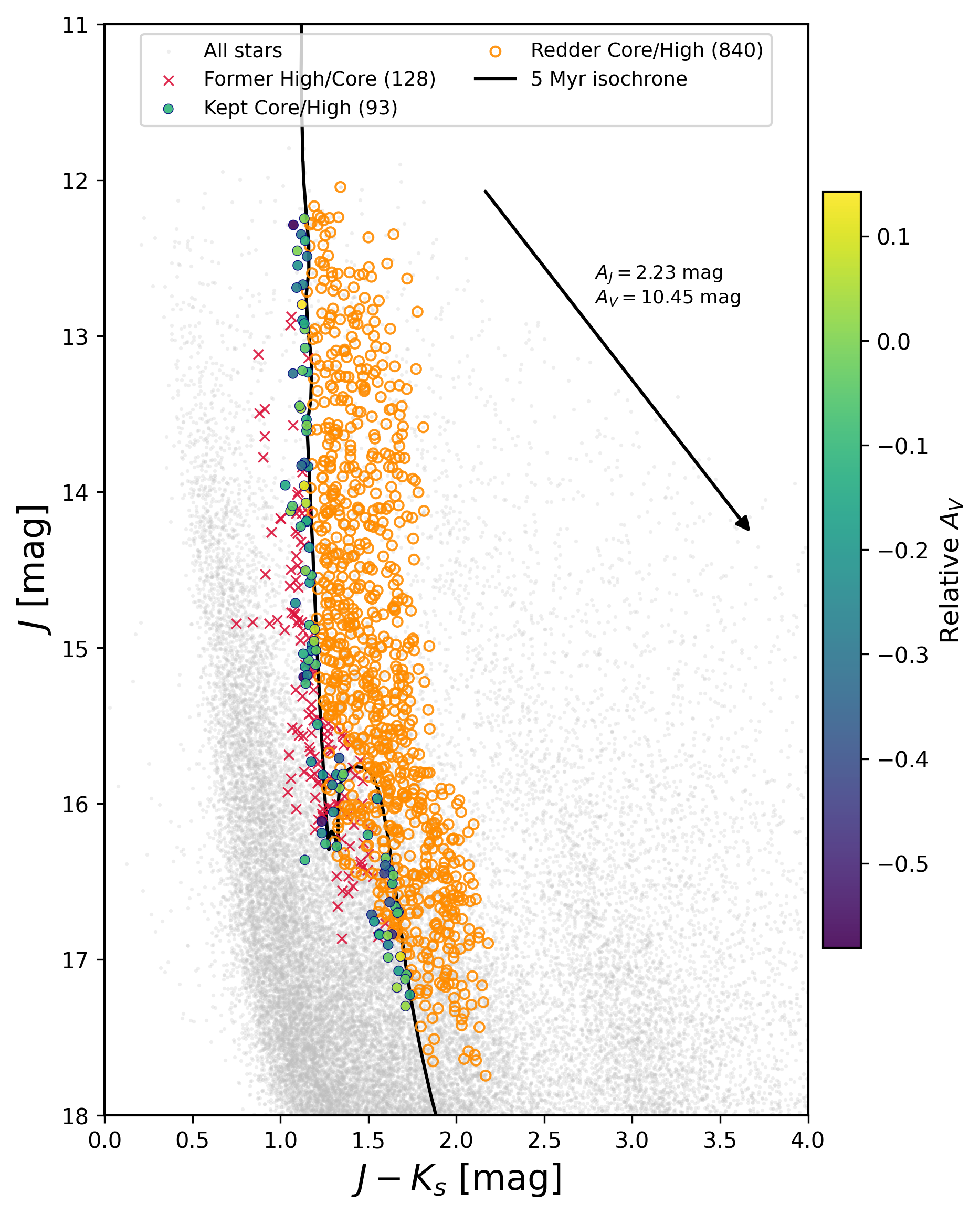}
   \caption{$J$ vs. $J-K_{\rm{s}}$ CMD for the refinement selection process. Core and high confidence stars redder than the $5\,\rm{Myr}$ isochrone (black line) are shown in dark orange open circles. The isochrone is located at a distance of $4.23\,\rm{kpc}$ and using an extinction value of $A_{V}=9\,\rm{mag}$, which is the lowest value expected for Wd1 members. Stars bluer than the isochrone, even considering their photometric errors, were discarded and are shown as red crosses. On the contrary, sources bluer than the isochrone, but that could be located over it due to their errors or that are intrinsically redder are shown as circles colored by their relative $A_{V}$ on the color bar. The mean extinction vector, $A_{V}=10.45\,\rm{mag}$ (or $A_{J}=2.23\,\rm{mag}$), for Wd1 members is also shown.}
    \label{refinement_example}%
\end{figure}

A small fraction of selected candidate members are bluer than both $5$ and $6$ Myr isochrones in the CMD, in a region populated by foreground stars. In order to limit contamination from the foreground, we retained only sources bluer than the isochrone, whose distance along the extinction vector from the isochrone is smaller than $2\sigma$. These retained sources are shown as circles in the example CMD of Fig.~\ref{refinement_example}, colored according to their extinction $A_V$ relative to the expected extinction of the isochrone. Discarded blue sources were around $10\%$ of the initial core$+$high confidence selection in both age cases.\\

\begin{figure}
    \centering
    \includegraphics[width=.4\textwidth]{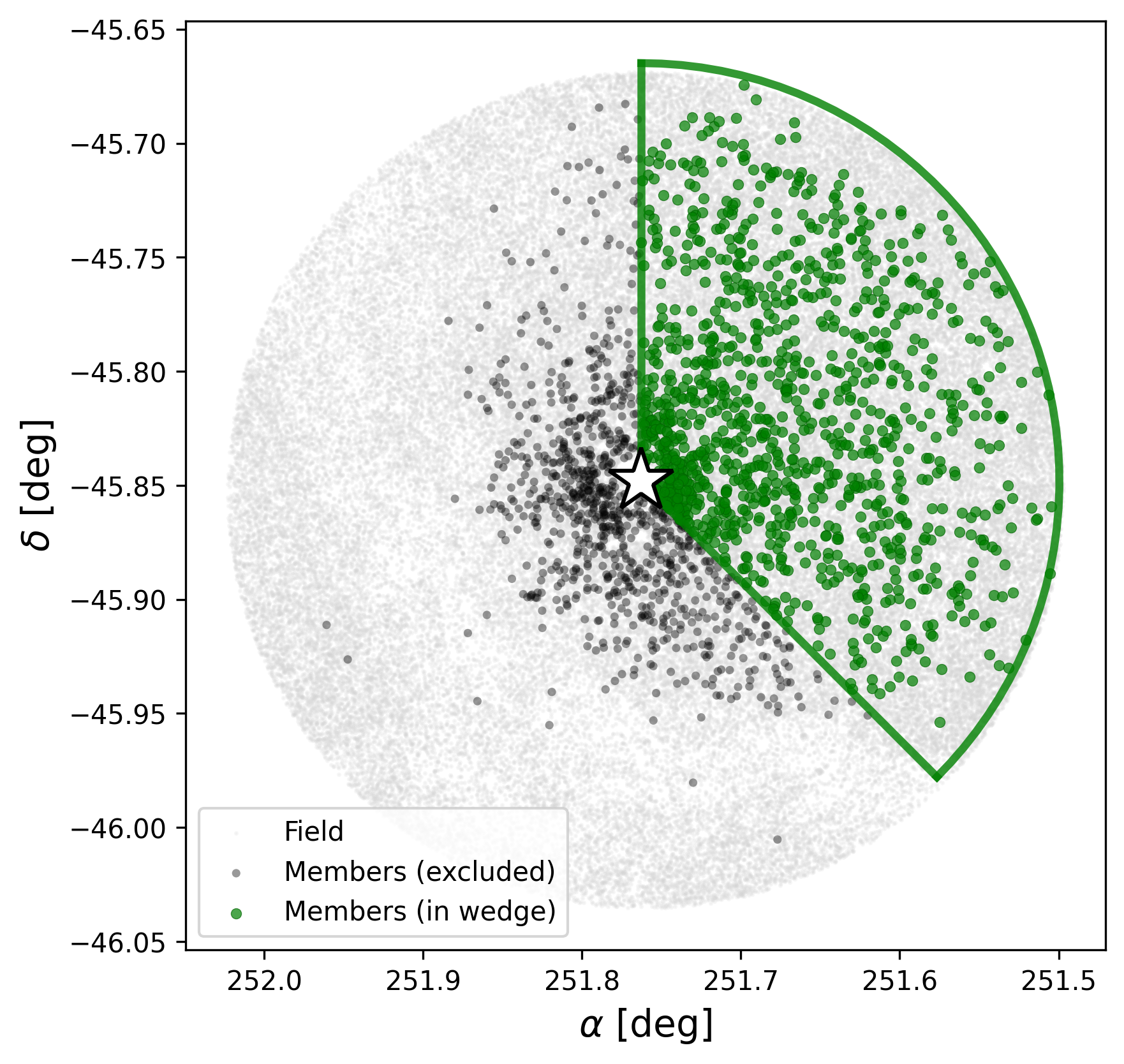}
   \caption{Spatial distribution of all stars in Wd1 region (in gray) and candidate members (in black) for $R_{s}=11\,\rm{arcmin}$, which was considered to compute the best wedge located in PA $225\degree$--$360 \degree$ associated to a low extinction sector (encircled in green) in Wd1 region. Green stars are within this best wedge and considered to compute the fraction of cluster members and their PM dispersion.}
    \label{wedge}%
\end{figure}

In order to find an optimal search radius that maximizes the selection of real members, minimizing spurious selection \citep{Sanchez_2010}, we computed the fraction of cluster members $N_{c}/N_{\rm{total}}$. Here, $N_{c}$ is the refined number of candidate members (core $+$ high-confidence without discarded blue stars) and $N_{\rm{total}}$ is the total number of stars within the search radius. However, due to the heavy extinction in the Wd1 region, the fraction $N_{c}/N_{\rm{total}}$ can be biased to zones of fewer star counts unrelated to cluster membership. To mitigate this, we restricted the computation of $N_{c}/N_{\rm{total}}$ to a subregion that has comparable foreground extinction to the center of Wd1. To identify such a subregion, we divided the field around Wd1 into $24$ azimuthal sectors of $15\degree$ each. For each sector, we counted the total number of stars and flagged those with counts above the median of all $24$ sectors as high-density. We then identified the longest contiguous sequence of high-density sectors, that is, the largest uninterrupted arc of the field with consistently high star counts, which we take as the region least affected by extinction or the best wedge. This procedure was repeated independently for each search radius $R_s$, and the same optimal sector was recovered in all cases. For $R_{s}=11\,\rm{arcmin}$ (the mean of all values explored), this best wedge spans $PA=225$--$360\,\rm{degree}$ and is shown in green in Fig.~\ref{wedge}. The same wedge was obtained when using either the $5$ or $6\,\rm{Myr}$ isochrone.\\

\begin{figure}
    \centering
    \includegraphics[width=.4\textwidth]{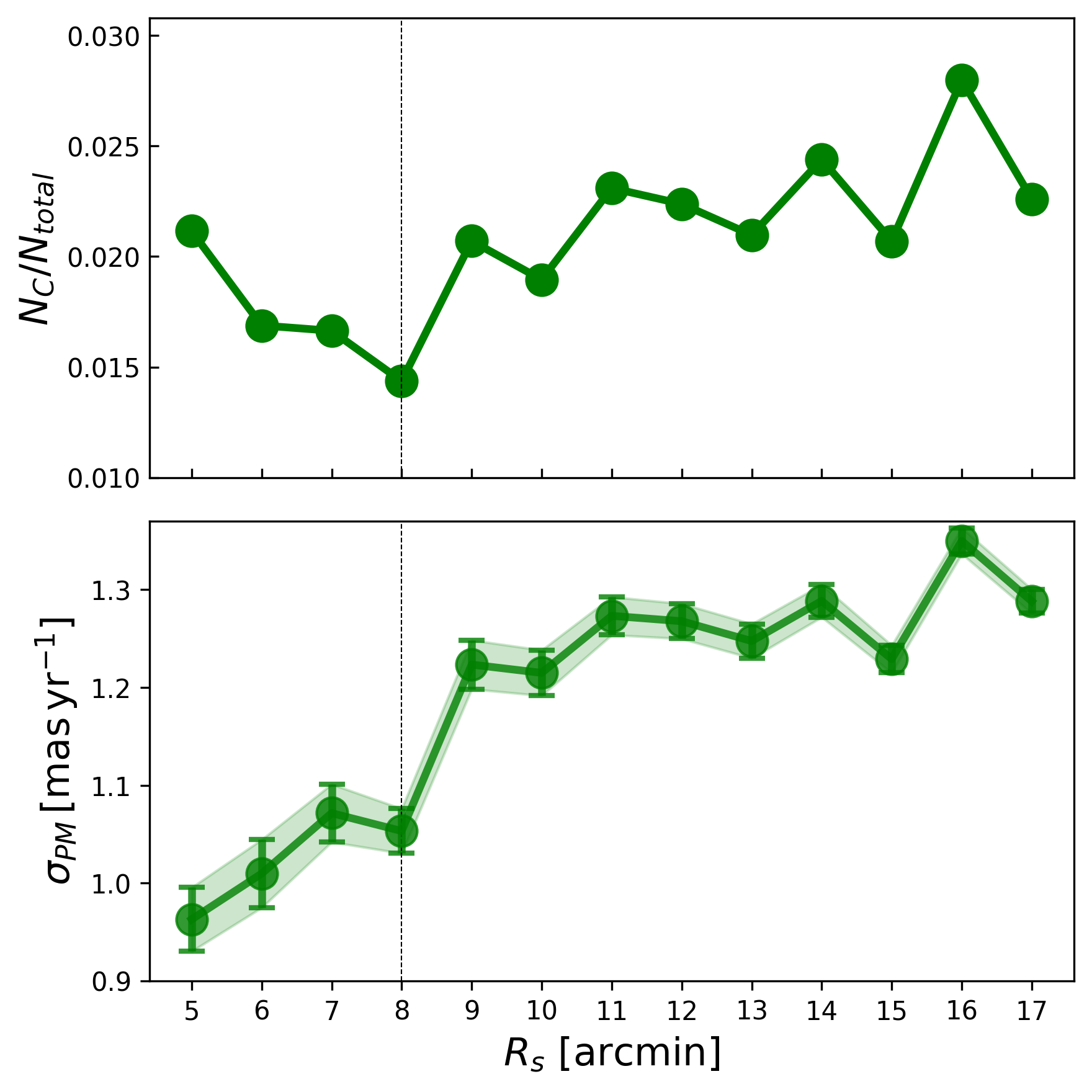}
   \caption{\textit{Upper panel:} Cluster member fraction against the search radius. The number of cluster members, $N_{C}$, and the total number of stars, $N_{\rm{total}}$, are computed only from the best wedge region described in the text. The vertical dashed line is located at $R_{s}=8\,\rm{arcmin}$, where the fraction stops decreasing. \textit{Lower panel:} As in the upper panel, but now showing the PM dispersion of identified cluster members for different values of $R_s$.}
    \label{members_wedge}%
\end{figure}

Stars inside this best wedge region are the ones considered to compute the fraction of cluster members against $R_{s}$. The upper panel of Fig.~\ref{members_wedge} shows how the fraction of cluster members decreases for $R_{s}\leq8\,\rm{arcmin}$, using a $5\,\rm{Myr}$ isochrone. This is the expected behavior for a cluster together with stars in the field. For higher $R_{s}$ values, the fraction of members starts to increase, which is an indication of spurious sources introduced by the algorithm \citep{Sanchez_2010}. This calculation also confirms that, for a $5\,\rm{Myr}$ isochrone, our optimal search radius should be fixed at $R_{s}=8\,\rm{arcmin}$, related to the minimum fraction of cluster members found.\\

To further confirm this optimal $R_{s}$ value, we also computed a combined PM dispersion $\sigma_{\mathrm{PM}} = \sqrt{\sigma^2_{\mu_\alpha \cos{\delta}} + \sigma^2_{\mu_\delta}}$ against the search radius. Without considering the effect of runaway stars, the PM dispersion of the selected candidates is not expected to increase as the search radius grows, as long as $R_s$ remains within the true cluster extent, since enlarging the aperture would only add more genuine members with similar kinematics. An increase in $\sigma_{\rm{PM}}$ for larger $R_s$ values therefore signals the growing inclusion of kinematically distinct field contaminants, confirming that the optimal radius has been exceeded. This is shown in the lower panel of Fig.~\ref{members_wedge}, where $\sigma_{\rm{PM}}$ errors have been calculated following \citet{Sanchez_2010} and using a bootstrap technique with $100$ random resamplings. In the plot, $R_{s} \geq 8\,\rm{arcmin}$ introduce a faster increase in $\sigma_{\rm{PM}}$, which again confirms our estimated optimal search radius $R_{s}=8\,\rm{arcmin}$. This radius includes $598$ core members and $335$ sources classified as high confidence, leaving us with $933$ member candidates of the Wd1 cluster. This optimal $R_s$ also indicates a possible larger size of Wd1 compared to the values in the literature, corresponding to a physical radius of $\sim 9.8\,\rm{pc}$, at a distance of $4.23\,\rm{kpc}$.\\

We repeated the analysis for candidate members found using the $6\,\rm{Myr}$ isochrone. In this case, the optimal radius $R_{s}$ was found to be $10\,\rm{arcmin}$ (corresponding to a physical size of $12.3\,\rm{pc}$). Nevertheless, since we are considering steps of $1\,\rm{arcmin}$, they are still in close agreement. Therefore, when using the $6\,\rm{Myr}$ isochrone, we considered candidates inside a $10\,\rm{arcmin}$ radius from the cluster center. This amounts to a total of $1\,162$ stars, with $752$ core and $410$ high confidence sources.\\

Note that stars on the best wedge are considered only to find the search radius $R_s$ through the cluster member fraction $N_{c}/N_{\rm{total}}$ and the PM dispersion $\sigma_{\rm{PM}}$. All the following analyses, but the surface density profile, are made considering the entire region and not only the sources within the best wedge zone. Thus, each isochrone has left us with two lists of candidates, which include $809$ common sources. To compile a final list of unique members, we combined these two lists and obtained $1286$ member candidates of Wd1. This number, depending on the IMF and total mass assumed, would correspond roughly to $10-30 \%$ of the total members of the cluster in the $1.5$--$20\,M_{\odot}$ mass range \citep{Lim_2013, Andersen_2017}.\\

Finally, we tested the sensitivity of the membership selection to the exclusion of the crowded central region. We find that the clustering results remain stable for inner exclusion radii up to $\sim 1\,\rm{arcmin}$. The mean PMs of the selected members change only marginally and the median PM uncertainties remain comparable to those obtained with smaller exclusions. For larger inner cuts ($\geq 2\,\rm{arcmin}$), however, the PM uncertainties increase and the separation between cluster and field populations becomes progressively less robust. This indicates that the high stellar density and spatial clustering in the central region play an important role in the {\tt HDBSCAN} algorithm and the recovery of high-probability members. Therefore, trying to maximize completeness over purity, in the following we present the results and our proposed census, obtained not applying any exclusion radius to the original VIRAC2 catalog.\\

\subsection{Population characteristics}
\label{candidates}

\begin{figure*}
    \centering
    \includegraphics[width=1.0\textwidth]{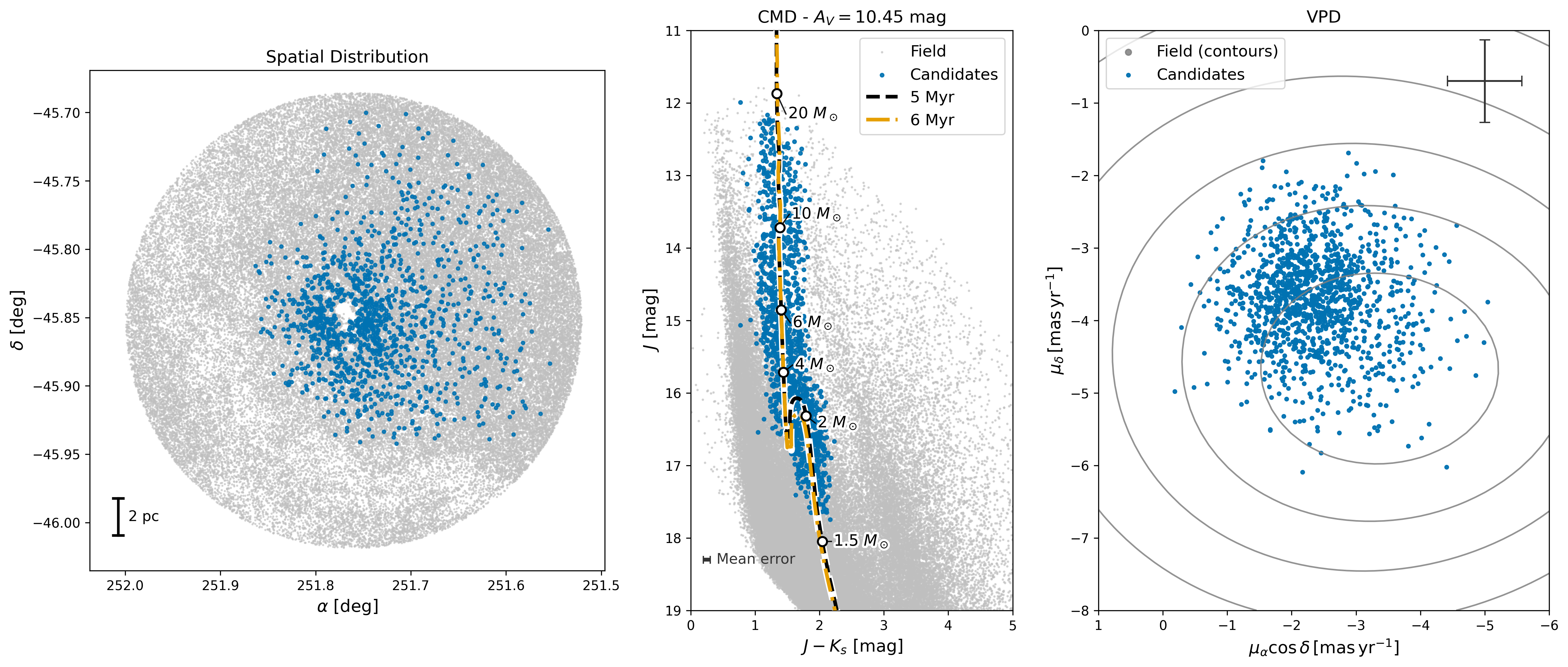}

   \caption{Final list of our Wd1 census (blue dots) compared to all field stars (gray) in $R_{s}=9\,\rm{arcmin}$. \textit{Left panel:} Spatial distribution of the candidate members. Scale length, according to the distance of the cluster, is on the bottom left. \textit{Central panel:} $J$ vs. $J-K_{\rm{s}}$ CMD. The black dashed line represents the PARSEC $5\,\rm{Myr}$ isochrone assuming a distance of $4.23\,\rm{kpc}$ and reddened for $A_{V}=10.45\,\rm{mag}$. The same assumptions were considered for the $6\,\rm{Myr}$ isochrone depicted with a dash-dotted orange line. For the $5\,\rm{Myr}$ isochrone, different mass values are also labeled. \textit{Right panel:} equatorial vector point diagram (VPD). Field stars are shown with gray contours and the median error for our selection of members is shown in the upper right.}
    \label{final_candidates}%
\end{figure*}

\subsubsection{Spatial distribution}

The spatial distribution of our proposed $1286$ Wd1 members is shown in the left panel of Fig.~\ref{final_candidates}. The latter shows that the members (blue dots) tend to be spherically concentrated in the central parts of the cluster with an extended halo to the north-east, potentially with additional structure. Field stars are colored in gray and distributed within $10\,\rm{arcmin}$ radius. To assess whether the stellar distribution is limited by extinction, we considered the dust emission map by \citet{Marsh_2017} in the same Wd1 region. We identified all VIRAC2 sources in this map and selected two equal circular regions: one in a high dust emission area and another in a region with four times lower dust emission. We found almost the same number of star counts in each, which suggests that VIRAC2 data is not affected by extinction, at least up to the distance of Wd1. Moreover, this also points to our stellar distribution being the real one for the cluster members.\\

The innermost part of the cluster, around $1\,\rm{arcmin}$ radius, contains fewer star counts compared to the outer regions. This is mainly because massive stars are located in this zone and the VVVX and VIRAC2 data are saturated for such bright stars. Saturation is strongest in the inner $\sim 1\,\rm{arcmin}$ and remains severe within $\sim 2\,\rm{arcmin}$, which motivates the $2\,\rm{arcmin}$ central mask adopted for the surface density fit. This observational choice should be distinguished from the membership selection robustness tests, which remain stable for inner exclusion radii up to $\sim 1\,\rm{arcmin}$ and degrade for larger cuts. On the other hand, in the south-west region an intense $A_{V}$ belt can be observed, translating into fewer star counts. The latter coincides with the cold dust emission region observed by \citet{Marsh_2017}. Our detected population is not located in this area.\\

\begin{figure}
    \centering
    \includegraphics[width=0.45\textwidth]{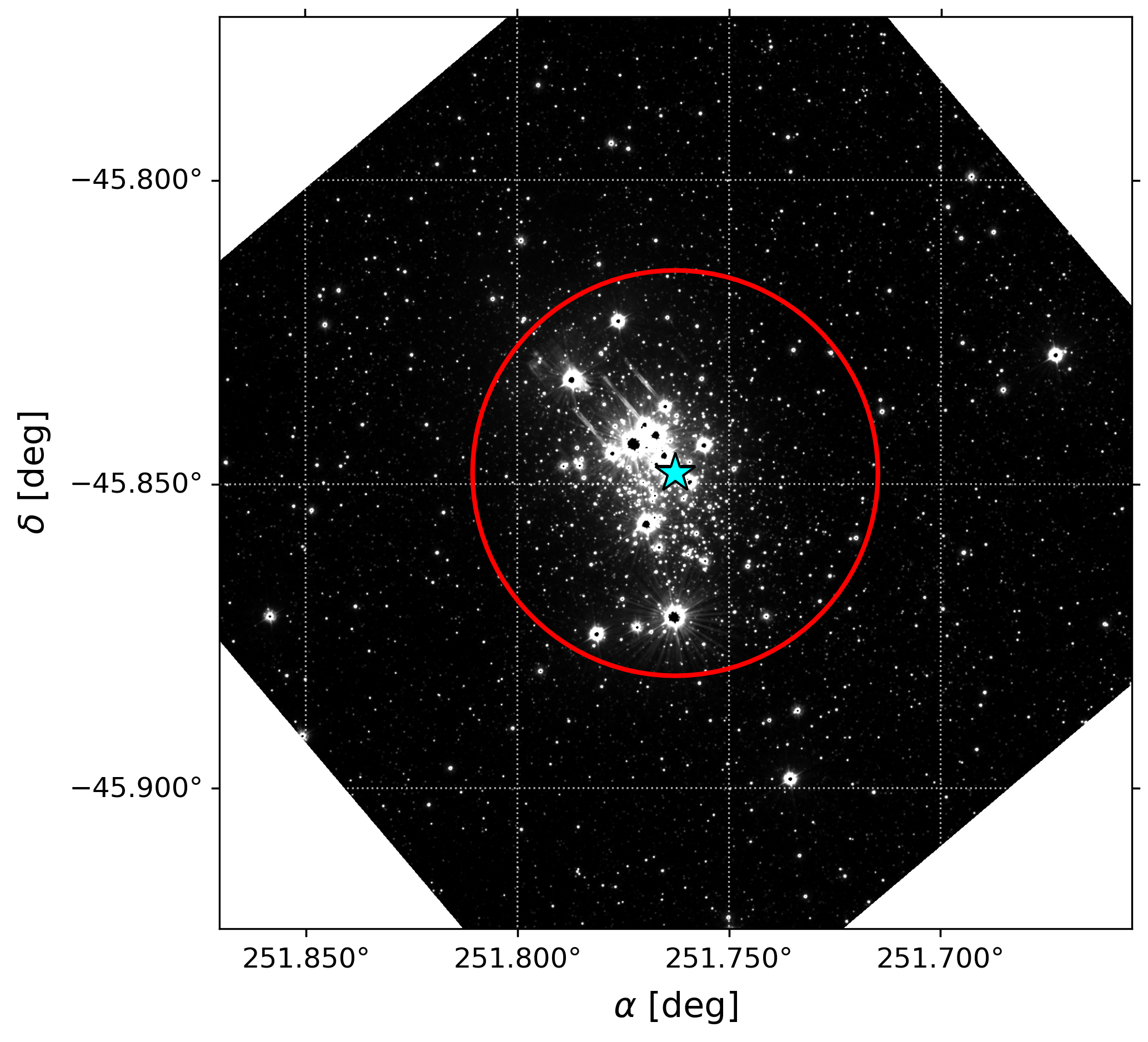}
   \caption{VVV $K_{\rm{s}}$ image of the $9\,\rm{arcmin} \times 9\,\rm{arcmin}$ region around Wd1. The cyan star shows the position of the center of the cluster considered in this study. The red circle depicts the exclusion region for the surface density fit of $2\,\rm{arcmin}$ radius. Central massive stars are clearly saturated in this band.}
    \label{VVV_Wd1}%
\end{figure}

Our list of members from the best wedge allows us to constrain the core radius of the cluster. We computed the surface density of our proposed members at different radii, but only considering $646$ sources located within the best wedge. In addition, since we have few star counts in the innermost parts of the cluster, we excluded the inner $2\,\rm{arcmin}$ circle around the center, leaving $460$ stars for the fit. Fig.~\ref{VVV_Wd1} shows a $9\,\rm{arcmin} \times 9\,\rm{arcmin}$ VVV $K_{\rm{s}}$ image of Wd1 region, where the saturation is evident inside the red circle of $2\,\rm{arcmin}$. At $4.23\,\rm{kpc}$, this corresponds to approximately $2.5\,\rm{pc}$. The central cyan star marks the position of the center of the cluster considered in this study.\\

The core and tidal radii were computed using a King profile and fitting it to the surface density of our members. This analysis is intended only to provide an approximate structural characterization, since Wd1 is too young to be in virial equilibrium \citep{Cottaar_2012, Wei_2025}. In addition, the fit excludes the inner $2\,\rm{arcmin}$ because the central region is strongly affected by saturation from the brightest cluster members in the VVVX/VIRAC2 data. Our exclusion tests show that the membership selection remains stable when masking up to $\sim 1\,\rm{arcmin}$, but degrades for larger inner cuts. Therefore, the $2\,\rm{arcmin}$ exclusion adopted for this fit should be regarded as a practical choice imposed by saturation, not as an optimal regime for the clustering analysis. The fit is constrained mainly by the outer envelope of the distribution and the obtained core radius should be regarded as an upper limit rather than a robust measurement. Further, when comparing our surface density distribution with the one found by \citet{Guarcello_2024} using X-ray data, there is a significant difference in the shape of the distribution, driven by the lack of usable star counts in the center of the cluster in the VIRAC2 data. This does not affect the surface density found through X-ray data, which rises steeply towards the center of Wd1.\\

\begin{figure}
    \centering
    \includegraphics[width=0.5\textwidth]{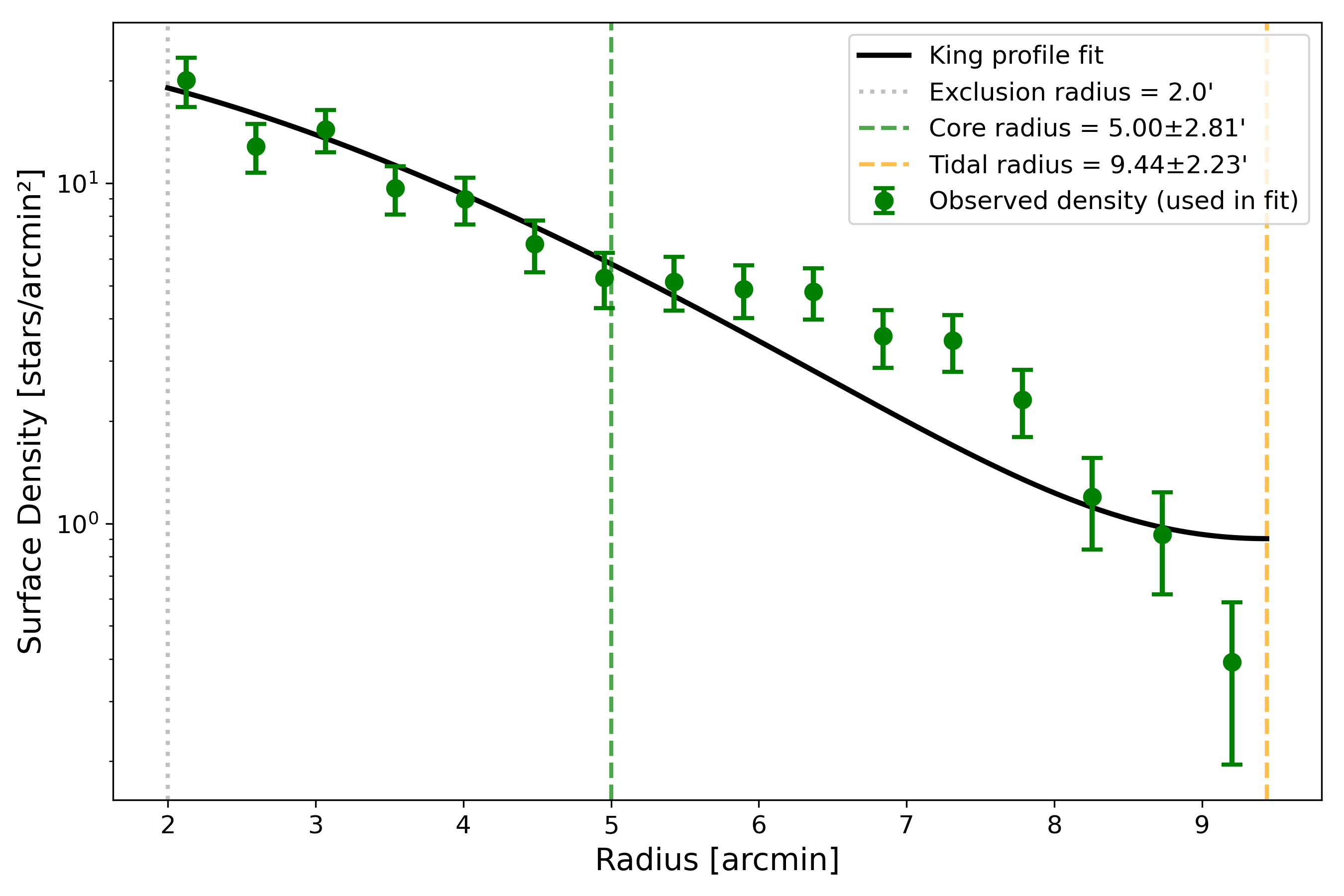}
   \caption{Surface density profile and their Poisson errors of $292$ of our proposed members located $2\,\rm{arcmin}$ or more away from the center of Wd1 and within the best wedge. The green dashed line indicates the computed core radius, the orange is related to the tidal radius and gray dotted line is the exclusion radius.}
    \label{density_king}%
\end{figure}

The surface density is shown in Fig.~\ref{density_king}, where the obtained core radius is about $5 \pm 2.81\,\rm{arcmin}$ (or $6.2 \pm 3.5\,\rm{pc}$) and the tidal radius is $9.44 \pm 2.23\,\rm{arcmin}$. Since we are not using the central parts of the cluster for the fitting, our result represents only an upper limit. The inner $2\,\rm{arcmin}$ exclusion zone, imposed by saturation of the most massive cluster members, encompasses the region where the King profile is most sensitive to the core radius. Thus, our fit is constrained by the outer envelope of the distribution. The derived core radius should be regarded as poorly constrained rather than a robust measurement.\\

Nevertheless, the entire distribution of members suggests a larger size for the Wd1 cluster, compared to the $2.5\,\rm{arcmin}$ radius measured by \citet{Lim_2013} or the $3.5\,\rm{arcmin}$ semi-major axis given by \citet{Negueruela_2022}. However, in the latter, the authors found a halo of redder, possible cluster members located up to $10\,\rm{arcmin}$ from the center. This means that the population in the outskirts of the distribution could be part of this halo of members. Our measured core radius value is larger than typical core radii for young massive clusters \citep{Santos-Silva_2012}, which should be more concentrated, consistent with it being an upper limit driven by the lack of data in the inner $2\,\rm{arcmin}$.\\

\subsubsection{NIR CMD}

The central panel of Fig.~\ref{final_candidates} shows the $J$ vs. $J-K_{\rm{s}}$ CMD of our members. The $5$ and $6\,\rm{Myr}$ PARSEC isochrones that we considered as input for the \textit{min\_dist} {\tt HDBSCAN} parameter are also shown as a dashed black line and a dotted-dashed line, respectively. Among our census, we selected sources with magnitudes $12\lesssim  J \lesssim 18\,\rm{mag}$, which, according to the isochrone, would have a mass range of $1.6$--$20\,M_{\odot}$ with the low-mass end located in the pre-Main Sequence phase. \\

This is deeper than the value reached by \citet{Gennaro_2017}, in which they also study members using NIR data, but with $50\%$ completeness down to $K_{s}=14.5\,\rm{mag}$ and in a much smaller region. Using VIRAC2, we can access lower mass stars because of the greater completeness of the overall catalog, which is $90\%$ for $K_{s}< 16\,\rm{mag}$ \citep{Smith_2025}. Most importantly, our list fills the gap between the massive members identified by \textit{Gaia} and very low-mass sources identified from EWOCS JWST data \citep{Guarcello_2025}, and where most of the X-ray counterparts are expected to be found. Although the three datasets are broadly complementary in the mass regimes they probe most completely, there is likely significant overlap, in particular, the \textit{Gaia}-based sample of \citet{Negueruela_2022} reaches down to $\sim 15\,M_{\odot}$ and the spectroscopic sample of \citet{Clark_2020} extends to at least $\sim 20\,M_{\odot}$, overlapping with the upper end of our VIRAC2 census. All samples are significantly incomplete, and the boundaries between them should be understood as indicative rather than sharp.\\

For a subset of our candidate members with available \textit{Gaia} DR3 astrometry, the mean PMs are consistent within uncertainties with previous determinations for Wd1. This agreement supports the reliability of our membership selection. A detailed kinematic analysis is presented in a companion paper (Ordenes-Huanca et al. 2026, in prep.), where we also introduce comparisons with recently published kinematic studies, such as the one by \citet{Wei_2025}.\\

\subsection{Variability analysis}
\label{variability}

In pre-Main Sequence stars, variability has different origins. Cold spots due to intense magnetic activity, presence of disks and variable accretion are the main causes of the observed changes. These variations have been observed in all wavelengths, including NIR (see, e.g. \citeauthor{Carpenter_2001}, \citeyear{Carpenter_2001}; \citeauthor{OrdenesHuanca_2022}, \citeyear{OrdenesHuanca_2022}). Open clusters are expected to host variable stars among their members \citep{Anderson_2025}. Main Sequence stars can be associated with pulsations inducing flux changes. The fraction of these variable sources is expected to increase for younger clusters and the physical origins of this behavior seem to diversify more at these ages. For a cluster of the age of Wd1, the main sources of variability are related to young stellar objects (YSOs) with discs or dark spots, as well as eclipsing binaries and pulsating stars. The latter can include $\alpha^{02}$ Canum Venaticorum stars and $\beta$ Cepheid variables. Moreover, it is expected that $50\%$ of members of a $5\,\rm{Myr}$ cluster should show signs of variability, at least in the optical regime. \\

\subsubsection{Excess variance}

The VIRAC2 catalog includes flux change indexes. Nevertheless, VVVX light curves are very sparse, unevenly sampled and with varying quality. Considering that and in order to analyze the variability signals in our census in a more robust way, we computed the excess variance metric $\sigma_{XS}^{2}$ \citep{Yuk_2022}. It is designed to measure variability that light curves show beyond their photometric uncertainties and is defined as

\begin{equation}
    \sigma_{XS}^{2}=S^{2}-\overline{\sigma_{\rm{error}}^{2}},
\end{equation}
\noindent
where $S^{2}$ is the variance of the light curve data points and $\overline{\sigma_{\rm{error}}^{2}}$ represents the mean square measurement error. Positive values of $\sigma_{XS}^{2}$ are expected to involve true variations, while negative values are variations only due to uncertainties. On the other hand, one can assign an error to the measured excess, $\Delta(\sigma_{XS}^{2})$, by assuming that $\overline{\sigma_{\rm{error}}^{2}}$ is negligible and taking into account the total number of epochs in a given light curve $N$. This error will be

\begin{equation}
    \Delta(\sigma_{XS}^{2})=\sqrt{\frac{2}{N-1}}S^{2}.
\end{equation}

\begin{figure}
    \centering
    \includegraphics[width=0.49\textwidth]{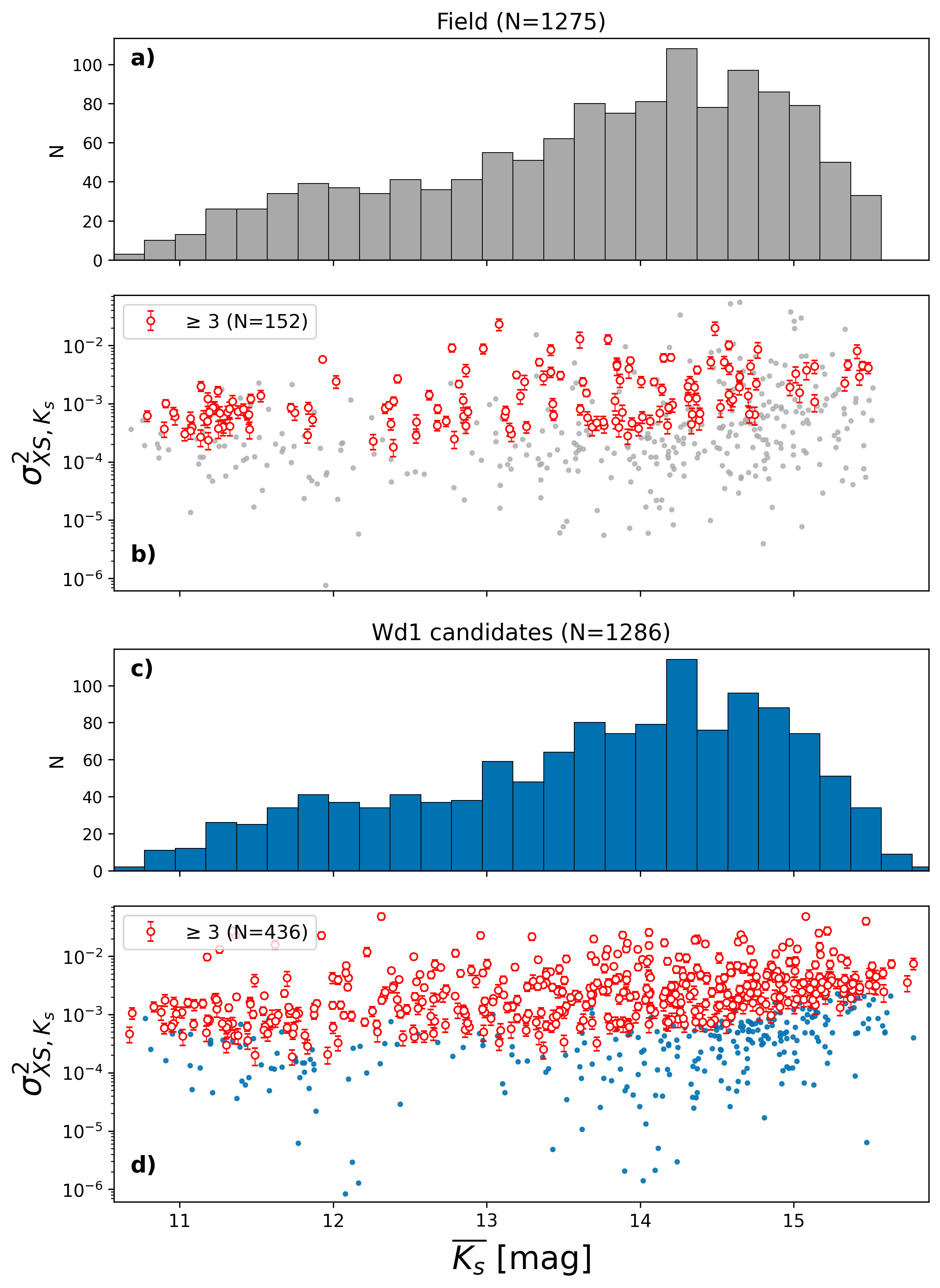}
   \caption{Excess variance $\sigma_{XS, K_{\rm{s}}}^{2}$ against the mean $K_{\rm{s}}$ magnitude of stars, $\overline{K_{\rm{s}}}$. Panel b) is for randomly selected field stars (in gray), whereas panel b) is for our $1286$ VIRAC2 candidate members of Wd1 (in blue). In both, objects showing real variations are highlighted in red open circles with their excess errors. Panels a) and c) show the $\overline{K_{\rm{s}}}$ distribution of field stars and Wd1 members, respectively.}
    \label{xs_comparison}%
\end{figure}

We measured these parameters for all our member candidates and plot their $\sigma_{XS, K_{\rm{s}}}^{2}$ against their mean $K_{\rm{s}}$ magnitudes in Fig~\ref{xs_comparison}. The calculation was carried out considering unsaturated data points on each light curve, with $chi < 5$ and $ast\_res\_chisq < 30$, as mentioned by \citet{Smith_2025}. In addition, we considered $ambiguous\_match = 0$ to avoid possible duplicated sources and $\textit{objtype}=1$, which indicates that sources are related to stars. \\

In order to investigate whether the observed variability is truly significant, we computed in the same way these variability metrics for $\sim 1\,200$ field stars randomly selected in the same $K_{\rm{s}}$ magnitude distribution of our candidates and located up to $10\,\rm{arcmin}$ from the center. According to \citet{Yuk_2022}, objects showing $\sigma_{\rm{XS}, K_{\rm{s}}}^{2}/\Delta\sigma_{\rm{XS}, K_{\rm{s}}}^{2} \geq 3.0$ are the ones presenting real flux changes. Therefore, in Fig~\ref{xs_comparison}, we highlighted in red open circles those that are above this threshold. In the left lower panel, field stars that meet this condition are only $154$, which represent about $12\%$ of the field sample. On the contrary, our proposed list of members (right lower panel of Fig~\ref{xs_comparison}), that meet the condition for real brightness changes, includes $34\%$ of the total number of sources. The $K_{\rm{s}}$ magnitude distribution for field and Wd1 members are shown in the upper left and upper right panels of Fig~\ref{xs_comparison}, respectively.\\

\begin{figure}
    \centering
    \includegraphics[width=0.5\textwidth]{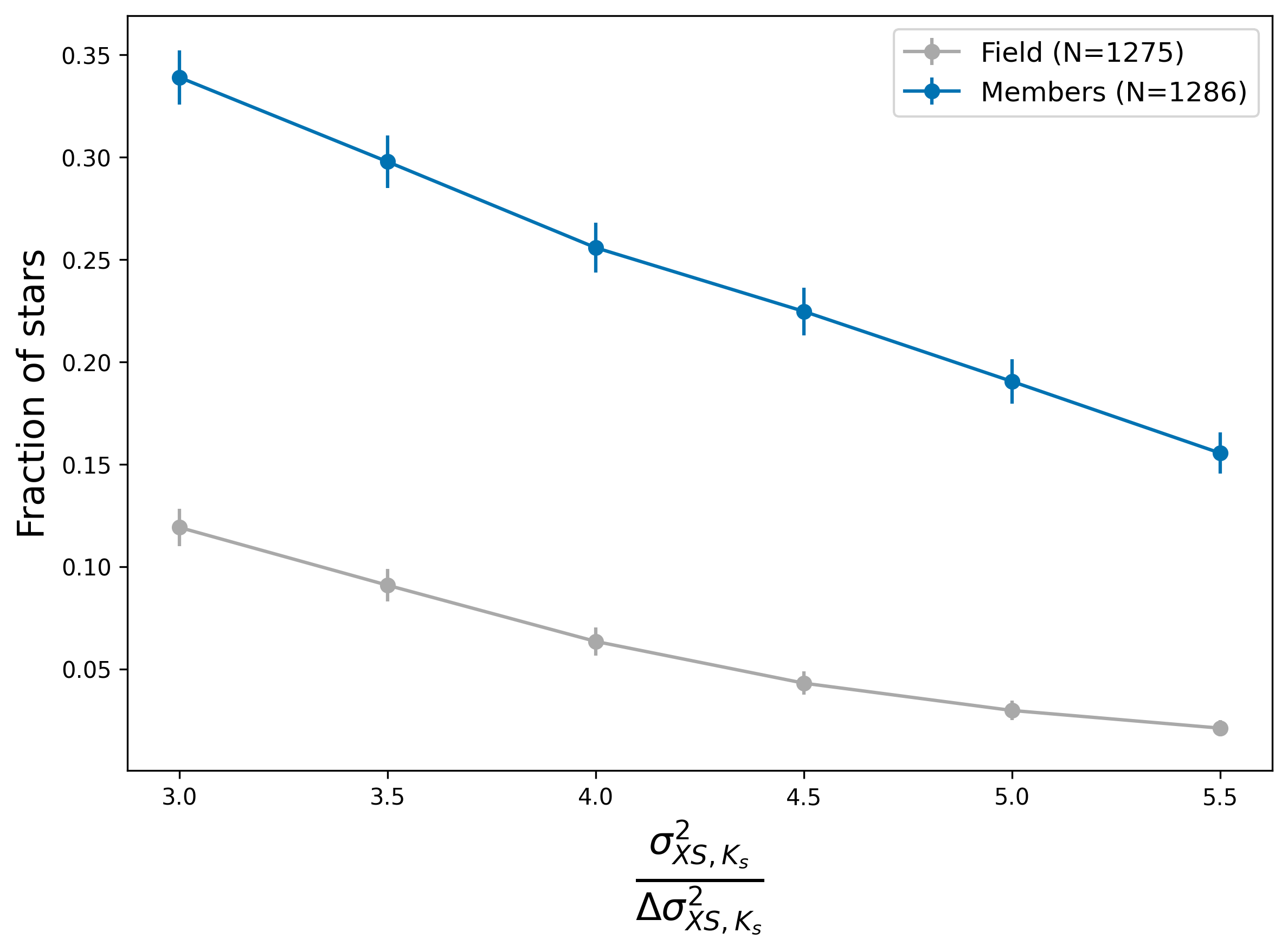}
   \caption{Fraction of variable stars in our proposed list of members (blue) compared to field stars (dark gray) against different threshold values to find variable behaviors. Binomial errors are also shown for each computation.}
    \label{xs_limit}%
\end{figure}

To confirm the variable behavior found in our sample, we increased the threshold to identify variable sources and computed the fraction of stars that met each condition. This is shown in Fig.~\ref{xs_limit}, where the fraction of variable stars for a given ${\sigma_{XS, K_{\rm{s}}}^{2}}/{\Delta\sigma_{XS, K_{\rm{s}}}^{2}}$ value is shown, with binomial errors. Field stars are shown in dark gray, while our members are shown in blue. It can be observed that, for any imposed threshold to find variable stars, the fraction of this type among our proposed members is higher than that of field stars. This also points to our proposed members being indeed associated with real brightness changes in their light curves.\\

\subsubsection{Variability classes}

For $1261$ out of the $1286$ sources from our census, with available light curves, we computed their Lomb-Scargle \citep{Lomb1976, Scargle1982} periodogram. We then classified these stars according to the periodicity $Q$ and asymmetry $M$ parameters defined by \citet{Cody_2014}. These were adapted to work on VVVX light curves by \citet{OrdenesHuanca_2022} and are extensively described there. They were mainly considered for pre-Main Sequence light curves, but they can also serve to identify other flux variations, such  as those possibly related to a companion. Summarizing them briefly here, the asymmetry $M$ parameter determines wether there are sudden increases or decreases in the observed flux, if any, and the periodicity $Q$ tells if the variations are periodic or not. One can measure these metrics for good quality light curves, which are those with at least $50$ epochs and mean photometric errors less than $20\%$ of the amplitude of variability, $\Delta K_{\rm{s}} = K_{\rm{s,max}}-K_{\rm{s,min}}$. From our $1286$ candidates, $846$ presented light curves that fulfill these conditions. These have a mean of $N=155$ epochs each, without considering data points $3\sigma$ away from the mean $K_{\rm{s}}$ magnitude of each.\\

\begin{figure}
    \centering
    \includegraphics[width=0.4\textwidth]{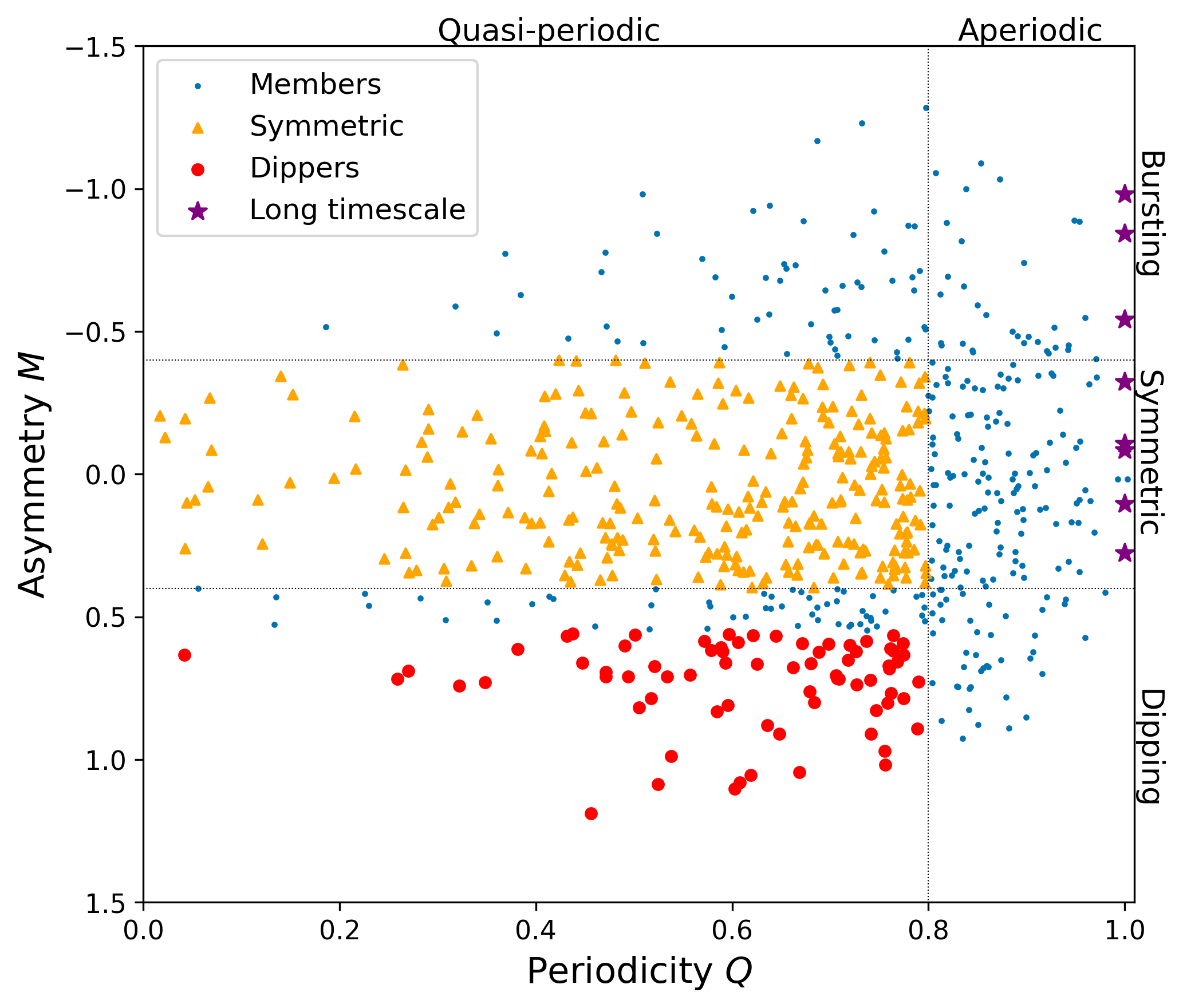}
   \caption{$Q$ vs. $M$ plane for all candidates with good quality light curves. Each region of the plot defines a variability class. The ones described in the text are highlighted using different colors and markers.}
    \label{QM}%
\end{figure}

The $Q$ metric usually presents values in the range $0\leq Q \leq 1$ and sources with stable periods are those with $Q\leq 0.6$. On the other hand, the asymmetry $M$ can adopt both positive and negative values. Negative (positive) values are related to increasing (decreasing) flux variations. Light curves with no net increase or decrease in brightness are usually in the range of $-0.4 \leq M \leq 0.4$ in VVVX data. Fig.~\ref{QM} shows the $Q$--$M$ plane, with all measured metric values for all the described variability classes. The combination of both metrics leads to $6$ different light curve classes, which are summarized in Table~\ref{tab:var_classes} and described in detail below.\\

\begin{table*}[h]
\caption{Summary of variability classes identified among Wd1 member candidates and discussed in the text.}
\label{tab:var_classes}
\centering
\begin{tabular}{lccc}
\hline\hline
Class & $Q$\,--\,$M$ criteria & Likely physical origin & $N$ \\
\hline
Quasi-periodic symmetric        & $Q < 0.8$,\ $-0.4 \leq M \leq 0.4$  & Spot modulation / rotation     & 265 \\
Periodic dipper  & $Q < 0.8$,\ $M > 0.55$               & Binaries / warped disks        & 44$^{a}$ \\
Long timescale   & $Q = 1$ (set), timescale $> 100$\,d  & Slow / unknown processes       & 8 \\
\hline
\end{tabular}
\tablefoot{$^{(a)}$\,Confirmed by visual inspection from an initial sample of 54 candidates.}
\end{table*}

The symmetric class is associated with periodic changes with no tendency for either flux increases or decreases. This is mainly related to dark spot rotational modulation. Using the $Q$ and $M$ metrics, we can identify periodic and nearly sinusoidal variations as the symmetric class of light curves, with $Q<0.8$ and $-0.4 \leq M \leq 0.4$. We found $265$ stars under this class, but the majority seem to have left their pre-Main Sequence stage given their $J$ magnitudes and their location on the isochrones. Despite the fact that young stars can be very faint and their photometric errors could be comparable to the amplitude of the variations, making it hard to detect a periodic signal, we could visually identify a few changes that seem to be periodic. We present two good examples of this class in Fig.~\ref{symmetric}, with $K_{\rm{s}} \geq 14.4\,\rm{mag}$ expected for pre-Main Sequence stars identified with a $5\,\rm{Myr}$ isochrone. Since our census also contains Main Sequence stars, another periodic brightness variation expected is from spotted solar-like stars with rotational modulation and pulsating $\delta$ Scuti stars.\\

\begin{figure*}
    \centering
    \includegraphics[width=0.7\textwidth]{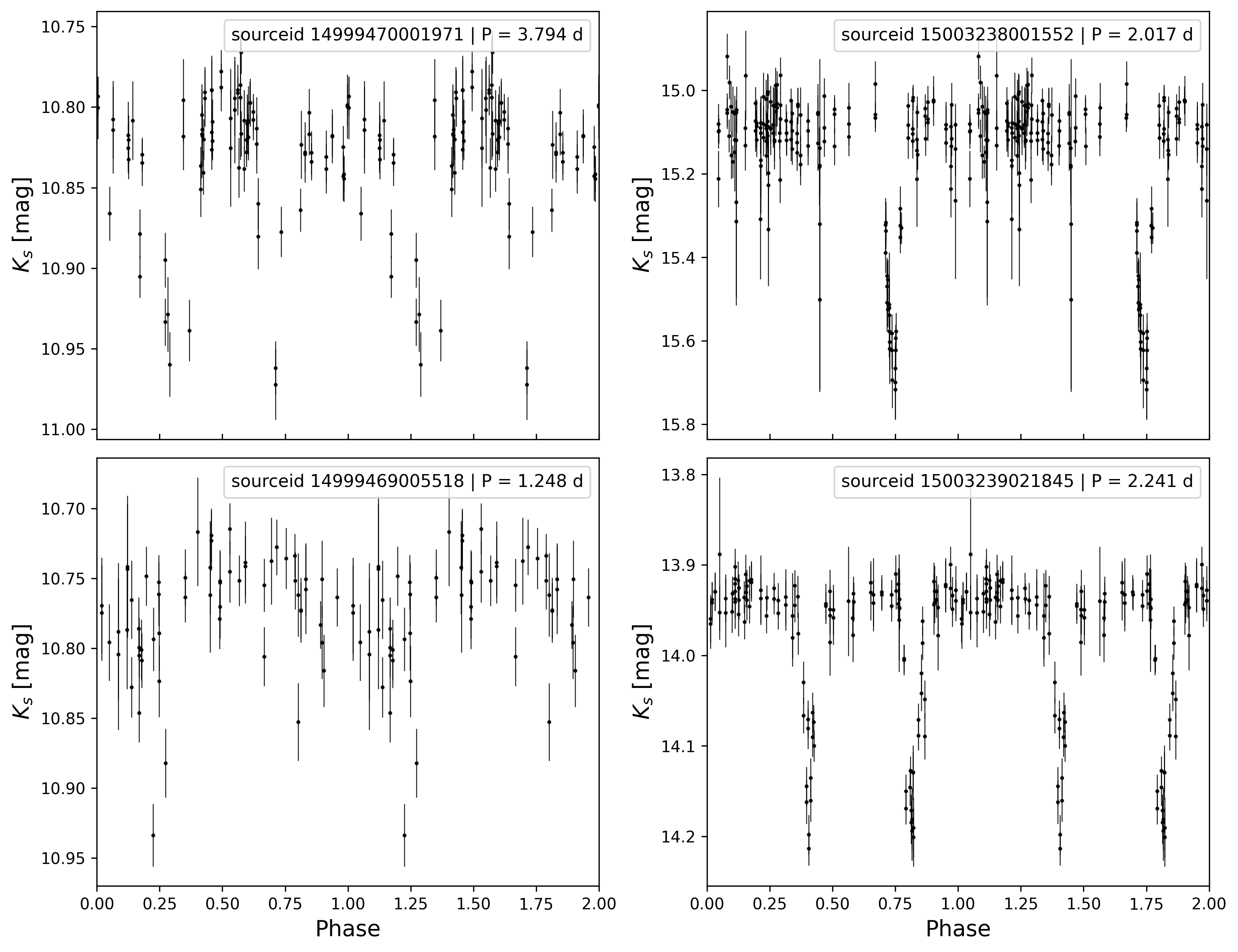}
   \caption{Examples of dipper $K_{\rm{s}}$ phase folded light curves found among member candidates. VIRAC2 source IDs are indicated in each panel along with their periods in days. The one in the upper right panel was already classified as an eclipsing binary EA/EB by \citet{Molnar_2022}.}
    \label{dippers}%
\end{figure*}

The periodic dipper category is for flux changes with a tendency to decrease periodically, which is mainly the case of binaries, but also for warped disks or substructures within them that transiently block the emission of a central star. Since the periodicity is computed using photometric errors, which could be higher in the fainter part of our dataset, we have relaxed the limit to identify periodic dipper light curves. At the same time, we imposed a higher limit for the asymmetry parameter to identify very strong and marked dips in the light curves. Then, periodic dipper objects will be the ones having $Q < 0.8$ and $M > 0.55$. Only $54$ of our proposed members met these conditions. We also visually inspected them to confirm the dipping changes. A total of $44$ dipper sources showed very marked, and possibly periodic, decreases in flux in their light curves, distinguishable from their uncertainties. Examples of them are shown in Fig.~\ref{dippers} with their VIRAC2 IDs and our computed periods. The one in the upper right panel has been classified as Algol-type or $\beta$ Lyrae-type (EA/EB) binaries by \citet{Molnar_2022} and those in the two bottom panels have already been identified as Wd1 members by \citet{Guarcello_2024}, showing X-ray emission. The one in the upper left panel is reported in the \citet{Cantat-Gaudin_2020} catalog of members. Since these sources are not hosting disks, according to their NIR colors, the observed dips can be due to a binary companion. All of them comprise our list of proposed binary candidates and their phase-folded light curves are provided in Appendix \ref{dippers_appendix}.\\

\begin{figure}
    \centering
    \includegraphics[width=0.35\textwidth]{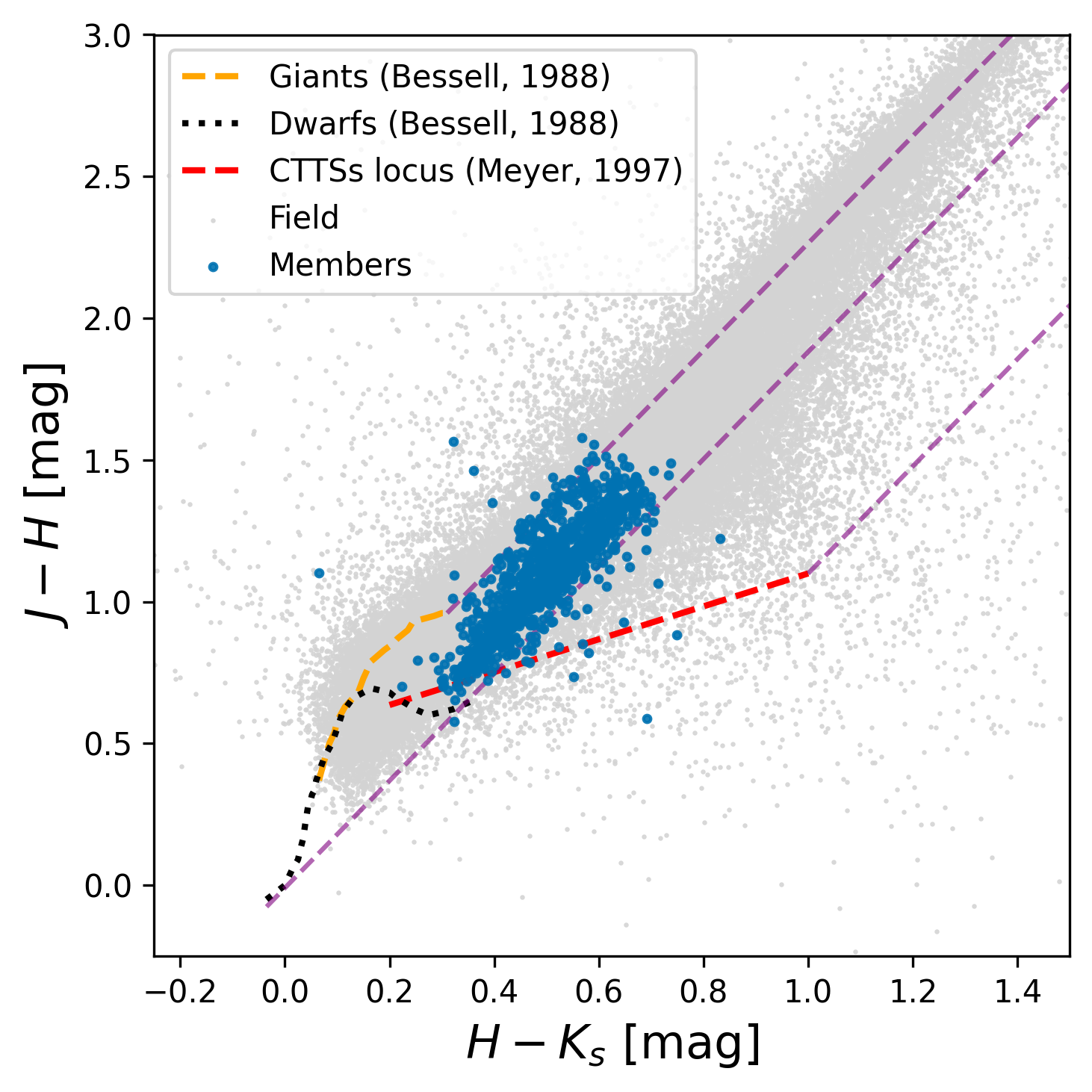}
   \caption{NIR color-color diagram $J-H$ vs. $H-K_{\rm{s}}$ of our candidates (blue dots) and field stars (gray dots). Colors expected for dwarfs and giants are depicted with black dotted and orange dashed lines, respectively. The CTTSs locus is a red dashed line, while purple lines show the direction of the extinction vector.}
    \label{CCD_candidates}%
\end{figure}

\begin{figure}
    \centering
    \includegraphics[width=0.45\textwidth]{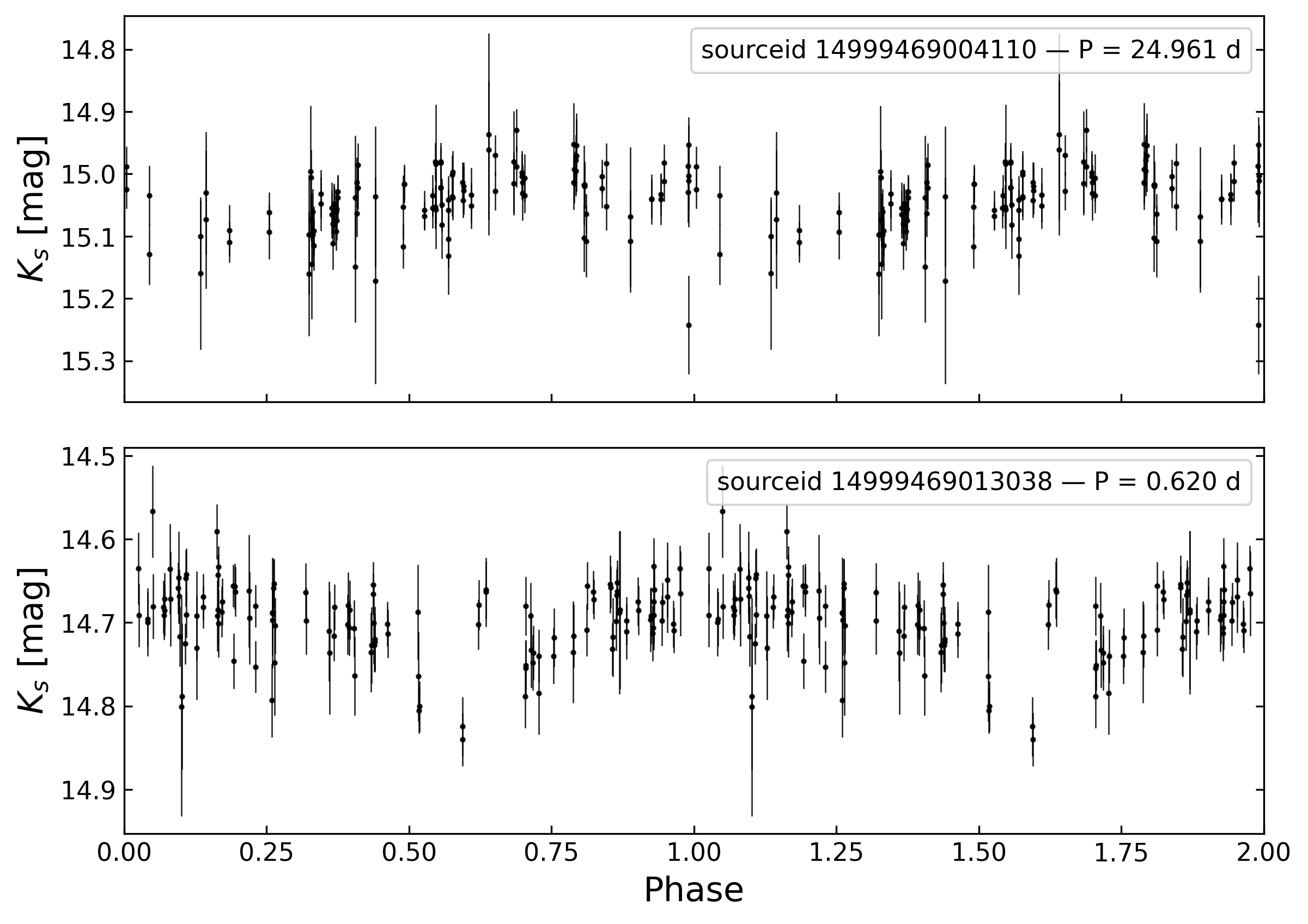}
   \caption{Examples of symmetric phase folded $K_{\rm{s}}$ light curves found in our proposed member candidates. VIRAC2 source IDs are indicated in each panel along with their periods. }
    \label{symmetric}%
\end{figure}

Those light curves showing dips with no clear repetitions are the ones called aperiodic dippers, while the ones with aperiodic flux increases are the ones called bursters. The latter two are related to the presence of disks and variable accretion, respectively \citep{Carpenter_2001, Stauffer_2016}. On the other hand, since we expect our list of candidates to contain few, if any, pre-Main Sequence stars (namely Classical T Tauri stars, CTTSs) with disks due to their proximity to the considered isochrone, dark spots should be the main source of flux variation in our stars. Thus, in Fig.\ref{CCD_candidates} we show the color-color diagram of our candidates (in blue), using the mean magnitudes reported in the VIRAC2 catalog, along with the locus of CTTSs as a dashed red line \citep{Meyer_1997}. Colors expected for dwarf and giants \citep{Bessell_1988} are shown as dotted black and orange lines, respectively. Also, lines indicating the direction of the extinction vector from \citet{Damineli_2016} are depicted by dashed gray lines. It is observed that our selection does not include pre-Main Sequence stars along this CTTSs locus, as expected by construction. Their deviation from dwarf and giant colors can be attributed only to extinction.\\

\begin{figure}
    \centering
    \includegraphics[width=0.45\textwidth]{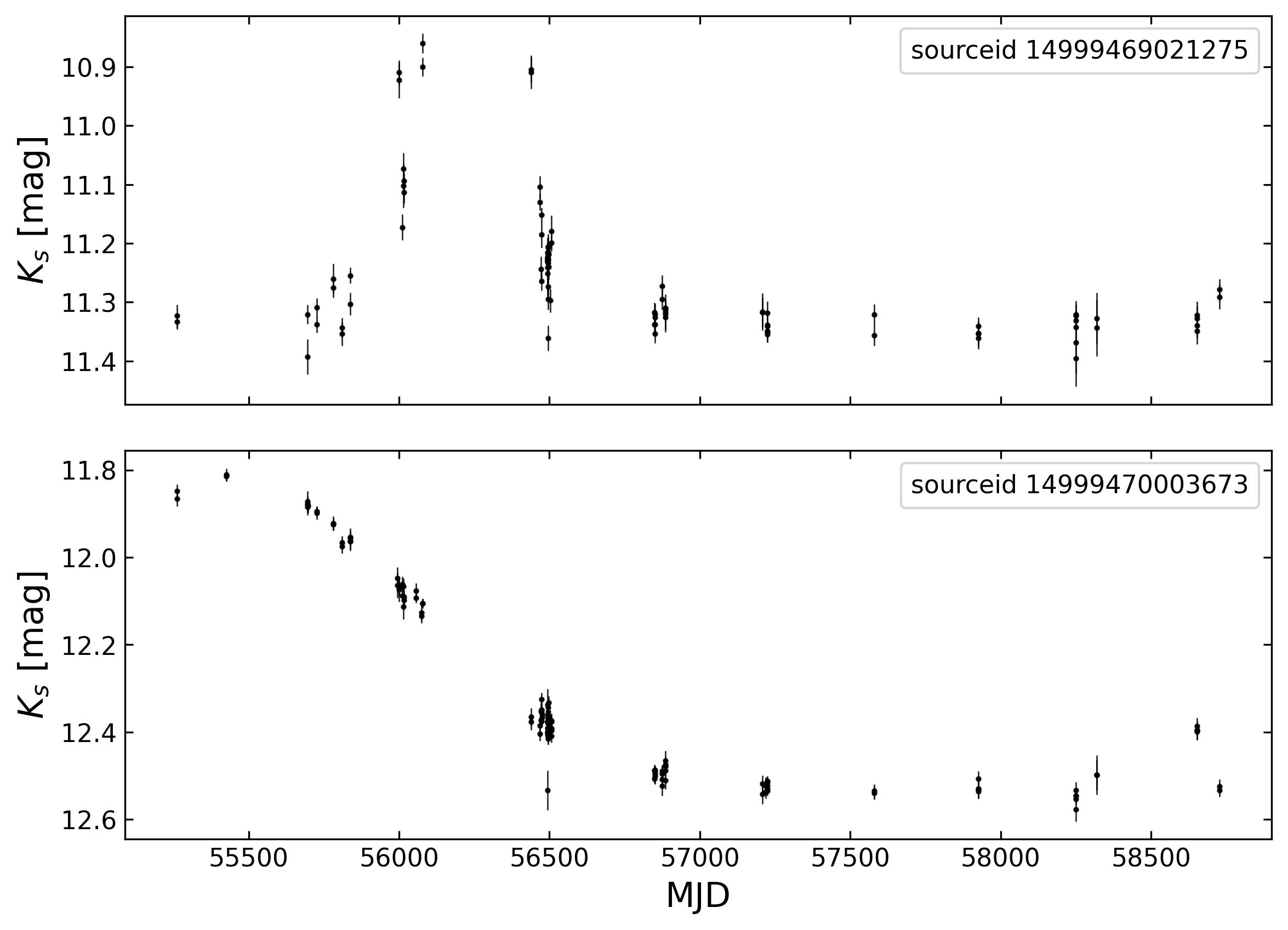}
   \caption{Examples of long timescale variables $K_{\rm{s}}$ light curves found in our proposed member candidates. VIRAC2 source IDs are indicated in each panel. Both have also been pointed as members of Wd1 by \citet{Gennaro_2017}.}
    \label{long}%
\end{figure}

\begin{figure}
    \centering
    \includegraphics[width=0.37\textwidth]{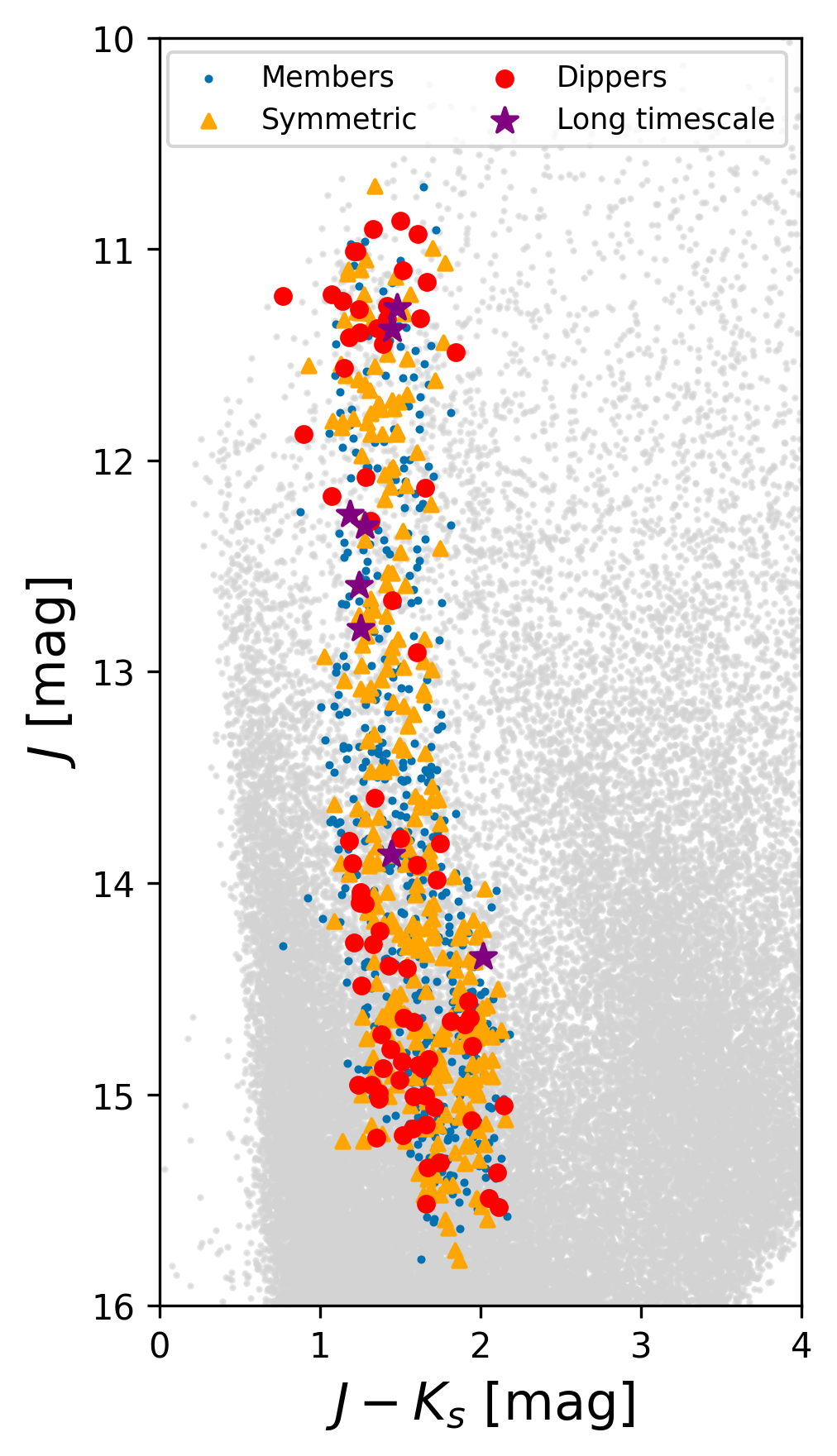}
   \caption{CMD of stars the region of Wd1 (in gray) against our variable candidates. The three tentative classes discussed in the text are highlighted with dippers in red circles, symmetric in orange triangles and long timescale sources as purple stars.}
    \label{cmd_qm}%
\end{figure}

Finally, stochastic flux changes describe sources showing no tendency to increase or decrease their brightness with no periodic changes. This class can be associated with different origins, such as variable accretion or the combination of several physical processes acting at the same time (see Fig. 8 in \citeauthor{Cody_2014} \citeyear{Cody_2014}). We have also included the long timescale variables for light curves with variation timescales larger than expected for variability or spot modulation \citep{Bonito_2023}. We set the periodicity value of long timescale variables to $Q = 1$. Additionally, we found $8$ sources with timescales of variation longer than $100$ days. Examples of this class are shown in Fig.~\ref{long}. Both sources have already been selected as Wd1 members by \citet{Gennaro_2017} and the one in the lower panel is also reported by \citet{Cantat-Gaudin_2020}. How these classes relate to the CMD of our candidates is shown in Fig.~\ref{cmd_qm}. Symmetric light curves distribute across all $J$ magnitudes, while dippers are preferentially located in the brighter parts of the CMD and the fainter region. The lack of this class for intermediate magnitudes is not clear.\\

\section{Discussion}
\label{discussion}

\subsection{Selection method comparison with literature}

Both isochrones considered in this work and the $J-K_{\rm{s}}$ color can limit us to find very reddened sources, since we keep the ones closer to it. The color-color diagram in Fig.~\ref{CCD_candidates} supports this statement. Previous studies have shown that the disk fraction in young clusters can be significantly affected by environmental effects such as external photoevaporation \citep{Balog_2007, Guarcello_2024}. Given Wd1's age and typical disk lifetimes, the fraction of sources with disks is expected to be low, with estimates suggesting around $20$--$30\%$ of members may still host a disk \citep{Michel_2021}. Furthermore, Wd1 presents an exceptionally extreme environment due to the presence of very high-mass stars, and we therefore expect the effects of photoevaporation on protoplanetary disk dispersal to be even more pronounced than in other young clusters, potentially reducing this fraction further. The detailed analysis of disk-hosting candidates and the search for evidence of external photoevaporation in Wd1 is the subject of ongoing work within the EWOCS project.\\

Also, the adopted extinction law from \citet{Damineli_2016} is assumed to be the one that is suitable for the cluster members. However, other extinction laws could affect our results. On the other hand, the combination of photometric and kinematic measurements can minimize false positives. \citet{Negueruela_2022} mentioned that PMs of Wd1 members can be indistinguishable from those of field stars and that photometric colors are a powerful tool to identify members, particularly on the Main Sequence. In our case, since we are using both, we can have a reliable list of member candidates by also restricting our findings to sources with very high probability of membership. Finally, the fraction of variable sources in our list of candidates is also pointing to a good selection. We think that the combination of all these metrics are leading to a comprehensive list of candidates and confirmed members of Wd1. \\

\subsection{Variable stars among Wd1 members}

High-mass binaries have been identified and studied within the cluster \citep{Koumpia_2012} and they represent a large fraction of the evolved members of Wd1 \citep{Ritchie_2022, Anastasopoulou_2025}. In the case of eclipses, the presence of a companion will decrease the observed flux, generating a dip over an almost constant level of flux. Our methodology proved to be effective in identifying binaries analyzed in the literature, but was also able to identify possible new, lower mass binary candidates. Due to the distance of Wd1, many of our sample of intermediate to low-mass stars are indeed faint. Thus, detecting real variations using ground-based data is challenging because in many cases these changes will be lost in the uncertainties of the photometric measurements. The number of variable stars found in this work can be a lower limit, given that there might still be a fraction of them unidentified through VIRAC2 data.\\

In addition, since we are using fixed isochrones to select stars near them, many of selected objects would not possess disks, leaving the observed flux changes mainly due to the presence of dark spots, binary interaction or pulsation for Main Sequence stars. The former should be the main source of variability on young open clusters \citep{Anderson_2025}. Still, its detection is even more difficult, since the temperature contrast between the spot and the photosphere of the star should be high enough to be detectable and translated into a periodic pattern in the light curve. This is why finding about $30\%$ of variable sources is very significant and can indicate that our selection is indeed related to young members of Wd1.\\

We are aware that the photometric errors can affect the $Q$ and $M$ metrics computation, but also the amplitude, cadence and sampling of VVVX and VIRAC2 data. Nevertheless, these data are very valuable since this is the first time we can have a glimpse of intermediate to low-mass light curves associated to Wd1 members. Further studies need to be made to confirm the proposed variability classes and better constrain the physical mechanisms at work.\\

\section{Summary}
\label{conclusions}

In this paper we used VIRAC2 data to find and effectively confirm $1286$ members of the super massive star cluster Westerlund~1. These objects can contribute to better constrain the general cluster properties of Wd1 by filling the gap in the mass range already covered by \textit{Gaia} on massive stars and JWST for low-mass sources \citep{Guarcello_2025}. \\

The members in our list were selected over an optimal $8'$ ($9.8\,\rm{pc}$) radius region around the cluster center, but preferentially towards its north-west region. This suggests that the size of Westerlund~1 is larger than often proposed in the literature. A surface density King profile \citep{King_1962, Tarricq_2022} allowed us to compute an upper limit for the size of the core radius of the cluster of $5.00\pm 2.81\,\rm{arcmin}$ ($6.2 \pm 3.5\,\rm{pc}$). This value is in agreement with previous results obtained from the analysis of smaller regions. Our candidates are distributed asymmetrically within the region defined by the search radius $R_s$ with a halo of members located towards the north-west (left panel on Fig.~\ref{final_candidates}). In order to verify the effects of heterogeneous extinction across the field, we performed the following test: Using the map by \citet{Marsh_2017}, we measured VIRAC2 star counts in a small region within this belt and in a four times higher extinction region with the same area in the same field of view, finding the same number of stars in each area. Having the same number of star counts in both suggests that the distribution that we have found is not only due to extinction but it may reflect a true asymmetry in cluster members distribution. We will investigate this further in an upcoming paper.\\

In addition to the photometric and astrometric selections, VVVX and VIRAC2 time series allowed to explore variability in candidate members and compare it with that of field stars. This is the first time that variability is studied in solar type stars of a young supermassive star cluster. Using the excess variance coefficient, about $30\%$ show signs of variability in the $K_{\rm{s}}$ band. Variability is an expected feature in this type of stars and support our method and selection independently.\\

To find different features of each variable light curve, we considered the $Q$ and $M$ parameters. This allowed us to find dips that can be possibly associated with binaries, nearly sinusoidal variations due to spot modulation, and flux changes on timescales longer than $100$ days. For possible binary related flux variations, we found $44$ sources showing visually confirmed dips in their light curves. These comprise our list of proposed binary candidates among Westerlund~1 members. Future EWOCS studies will explore the properties and the origin of different types of variability from a multi-wavelength approach.\\

A kinematic study of our proposed Westerlund~1 members will be presented in a companion paper (Ordenes-Huanca et al. submitted). This first part presents the first variability analysis of the intermediate-mass stellar content of the cluster, contributing to filling the gap between mass ranges.\\

\begin{acknowledgements}
      The authors thank the anonymous referee for the suggestions that improved this paper. This work has been supported by the Millenium Institute of Astrophysics MAS ICM-ANID, Comité Mixto ESO-Chile Postdoctoral Fellowship and the ANID BASAL Center for Astrophysics and Associated Technologies (CATA) through grant FB210003 AIM23-0001. COH acknowledges Agencia Nacional de Investigación y Desarrollo (ANID) through FONDECYT postdoctoral grant 3260854. AB acknowledges support under Germany’s Excellence Strategy through the Cluster of Excellence ORIGINS EXC–2094–390783311. RB acknowledges support by the INAF Mini-Grant “Physical properties of Accreting young stellar objects: exploration of their light Curves and Emission (PACE)”. KM acknowledges support from the Fundação para a Ciência e a Tecnologia (FCT) through the CEEC-individual contract 2022.03809.CEECIND, and grant UID/04434/2023. This research was supported by the International Space Science Institute (ISSI) in Bern, through ISSI International Team project \#25-639.
\end{acknowledgements}


\bibliography{Wd1_bibliography}

@ARTICLE{Tarricq_2022,
       author = {{Tarricq}, Y. and {Soubiran}, C. and {Casamiquela}, L. and {Castro-Ginard}, A. and {Olivares}, J. and {Miret-Roig}, N. and {Galli}, P.~A.~B.},
        title = "{Structural parameters of 389 local open clusters}",
      journal = {\aap},
         year = 2022,
        month = mar,
       volume = {659},
          eid = {A59},
        pages = {A59},
          doi = {10.1051/0004-6361/202142186},
archivePrefix = {arXiv},
       eprint = {2111.05291},
 primaryClass = {astro-ph.GA},
       adsurl = {https://ui.adsabs.harvard.edu/abs/2022A&A...659A..59T}
}

@ARTICLE{King_1962,
       author = {{King}, Ivan},
        title = "{The structure of star clusters. I. an empirical density law}",
      journal = {\aj},
         year = 1962,
        month = oct,
       volume = {67},
        pages = {471},
          doi = {10.1086/108756},
       adsurl = {https://ui.adsabs.harvard.edu/abs/1962AJ.....67..471K}
}

@ARTICLE{Castellanos_2026,
       author = {{Castellanos}, R. and {Najarro}, F. and {Garcia}, M. and {Negueruela}, I. and {Patrick}, L.~R. and {Ritchie}, B. and {Guarcello}, M.~G. and {Shenar}, T. and {Evans}, C. and {Prinja}, R. and {Fenech}, D.},
        title = "{First spectroscopic identification of the main sequence in Westerlund 1}",
      journal = {\aap},
         year = 2026,
        month = apr,
       volume = {708},
          eid = {A209},
        pages = {A209},
          doi = {10.1051/0004-6361/202558099},
archivePrefix = {arXiv},
       eprint = {2602.24218},
 primaryClass = {astro-ph.GA},
       adsurl = {https://ui.adsabs.harvard.edu/abs/2026A&A...708A.209C}
}

@ARTICLE{Ordenes-Huanca_2024,
       author = {{Ordenes-Huanca}, C. and {Zoccali}, M. and {Bayo}, A. and {Cuadra}, J. and {Contreras Ramos}, R. and {Rojas-Arriagada}, A.},
        title = "{Unveiling the structural content of NGC 6357 via kinematics and NIR variability}",
      journal = {\mnras},
         year = 2024,
        month = sep,
       volume = {533},
       number = {1},
        pages = {841-859},
          doi = {10.1093/mnras/stae1862},
archivePrefix = {arXiv},
       eprint = {2407.17577},
 primaryClass = {astro-ph.SR},
       adsurl = {https://ui.adsabs.harvard.edu/abs/2024MNRAS.533..841O}
}

@ARTICLE{Santos-Silva_2012,
       author = {{Santos-Silva}, T. and {Gregorio-Hetem}, J.},
        title = "{Characterisation of young stellar clusters}",
      journal = {\aap},
         year = 2012,
        month = nov,
       volume = {547},
          eid = {A107},
        pages = {A107},
          doi = {10.1051/0004-6361/201219695},
archivePrefix = {arXiv},
       eprint = {1209.1585},
 primaryClass = {astro-ph.GA},
       adsurl = {https://ui.adsabs.harvard.edu/abs/2012A&A...547A.107S}
}

@article{Balog_2007,
doi = {10.1086/513311},
url = {https://doi.org/10.1086/513311},
year = {2007},
month = {may},
publisher = {},
volume = {660},
number = {2},
pages = {1532},
author = {Balog, Zoltan and Muzerolle, James and Rieke, G. H. and Su, Kate Y. L. and Young, Eric T. and Megeath, S. Tom},
title = {Spitzer/IRAC-MIPS Survey of NGC 2244: Protostellar Disk Survival in the Vicinity of Hot Stars},
journal = {\apj}
}

@ARTICLE{Michel_2021,
       author = {{Michel}, Arnaud and {van der Marel}, Nienke and {Matthews}, Brenda C.},
        title = "{Bridging the Gap between Protoplanetary and Debris Disks: Separate Evolution of Millimeter and Micrometer-sized Dust}",
      journal = {\apj},
         year = 2021,
        month = nov,
       volume = {921},
       number = {1},
          eid = {72},
        pages = {72},
          doi = {10.3847/1538-4357/ac1bbb},
archivePrefix = {arXiv},
       eprint = {2104.05894},
 primaryClass = {astro-ph.EP},
       adsurl = {https://ui.adsabs.harvard.edu/abs/2021ApJ...921...72M}
}

@article{Cottaar_2012,
	author = {{Cottaar}, M. and {Meyer, M. R.} and {Andersen, M.} and {Espinoza, P.}},
	title = {Is the massive young cluster Westerlund I bound?},
	DOI= "10.1051/0004-6361/201117722",
	url= "https://doi.org/10.1051/0004-6361/201117722",
	journal = {A\&A},
	year = 2012,
	volume = 539,
	pages = "A5",
	month = "",
}

@ARTICLE{Wei_2025,
       author = {{Wei}, Lingfeng and {Boyle}, Peter C. and {Lu}, Jessica R. and {Hosek}, Jr., Matthew W. and {Konopacky}, Quinn M. and {Spencer}, Richard G. and {Kim}, Dongwon and {Rui}, Nicholas Z. and {Service}, Max and {Huang}, D.~B. and {Anderson}, Jay},
        title = "{Structure and Dynamics of the Young Massive Star Cluster Westerlund 1}",
      journal = {\apj},
         year = 2025,
        month = oct,
       volume = {992},
       number = {2},
          eid = {213},
        pages = {213},
          doi = {10.3847/1538-4357/adfc60},
archivePrefix = {arXiv},
       eprint = {2501.16755},
 primaryClass = {astro-ph.GA},
       adsurl = {https://ui.adsabs.harvard.edu/abs/2025ApJ...992..213W}
}

@ARTICLE{Beasor_2021,
       author = {{Beasor}, Emma R. and {Davies}, Ben and {Smith}, Nathan and {Gehrz}, Robert D. and {Figer}, Donald F.},
        title = "{The Age of Westerlund 1 Revisited}",
      journal = {\apj},
         year = 2021,
        month = may,
       volume = {912},
       number = {1},
          eid = {16},
        pages = {16},
          doi = {10.3847/1538-4357/abec44},
archivePrefix = {arXiv},
       eprint = {2103.02609},
 primaryClass = {astro-ph.SR},
       adsurl = {https://ui.adsabs.harvard.edu/abs/2021ApJ...912...16B}
}

@ARTICLE{Kudryavtseva_2012,
       author = {{Kudryavtseva}, Natalia and {Brandner}, Wolfgang and {Gennaro}, Mario and {Rochau}, Boyke and {Stolte}, Andrea and {Andersen}, Morten and {Da Rio}, Nicola and {Henning}, Thomas and {Tognelli}, Emanuele and {Hogg}, David and {Clark}, Simon and {Waters}, Rens},
        title = "{Instantaneous Starburst of the Massive Clusters Westerlund 1 and NGC 3603 YC}",
      journal = {\apjl},
         year = 2012,
        month = may,
       volume = {750},
       number = {2},
          eid = {L44},
        pages = {L44},
          doi = {10.1088/2041-8205/750/2/L44},
archivePrefix = {arXiv},
       eprint = {1204.5481},
 primaryClass = {astro-ph.GA},
       adsurl = {https://ui.adsabs.harvard.edu/abs/2012ApJ...750L..44K}
}

@ARTICLE{Stauffer_2016,
       author = {{Stauffer}, John and {Cody}, Ann Marie and {Rebull}, Luisa and {Hillenbrand}, Lynne A. and {Turner}, Neal J. and {Carpenter}, John and {Carey}, Sean and {Terebey}, Susan and {Morales-Calder{\'o}n}, Mar{\'\i}a and {Alencar}, Silvia H.~P. and {McGinnis}, Pauline and {Sousa}, Alana and {Bouvier}, Jerome and {Venuti}, Laura and {Hartmann}, Lee and {Calvet}, Nuria and {Micela}, Giusi and {Flaccomio}, Ettore and {Song}, Inseok and {Gutermuth}, Rob and {Barrado}, David and {Vrba}, Frederick J. and {Covey}, Kevin and {Herbst}, William and {Gillen}, Edward and {Medeiros Guimar{\~a}es}, Marcelo and {Bouy}, Herve and {Favata}, Fabio},
        title = "{CSI 2264: Characterizing Young Stars in NGC 2264 with Stochastically Varying Light Curves}",
      journal = {\aj},
         year = 2016,
        month = mar,
       volume = {151},
       number = {3},
          eid = {60},
        pages = {60},
          doi = {10.3847/0004-6256/151/3/60},
archivePrefix = {arXiv},
       eprint = {1601.03326},
 primaryClass = {astro-ph.SR},
       adsurl = {https://ui.adsabs.harvard.edu/abs/2016AJ....151...60S}
}

@ARTICLE{Bonito_2023,
       author = {{Bonito}, R. and {Venuti}, L. and {Ustamujic}, S. and {Yoachim}, P. and {Street}, R.~A. and {Prisinzano}, L. and {Hartigan}, P. and {Guarcello}, M.~G. and {Stassun}, K.~G. and {Giannini}, T. and {Feigelson}, E.~D. and {Caratti o Garatti}, A. and {Orlando}, S. and {Clarkson}, W.~I. and {McGehee}, P. and {Bellm}, E.~C. and {Gizis}, J.~E.},
        title = "{Young Stellar Objects, Accretion Disks, and Their Variability with Rubin Observatory LSST}",
      journal = {\apjs},
         year = 2023,
        month = mar,
       volume = {265},
       number = {1},
          eid = {27},
        pages = {27},
          doi = {10.3847/1538-4365/acb684},
archivePrefix = {arXiv},
       eprint = {2302.00897},
 primaryClass = {astro-ph.SR},
       adsurl = {https://ui.adsabs.harvard.edu/abs/2023ApJS..265...27B}
}

@article{Pedregosa_2011,
  title={Scikit-learn: Machine Learning in Python},
  author={Fabian Pedregosa and Ga{\"e}l Varoquaux and Alexandre Gramfort and Vincent Michel and Bertrand Thirion and Olivier Grisel and Mathieu Blondel and Gilles Louppe and Peter Prettenhofer and Ron Weiss and Ron J. Weiss and J. Vanderplas and Alexandre Passos and David Cournapeau and Matthieu Brucher and Matthieu Perrot and E. Duchesnay},
  journal={J. Mach. Learn. Res.},
  year={2011},
  volume={12},
  pages={2825-2830},
  url={https://api.semanticscholar.org/CorpusID:10659969}
}

@ARTICLE{Bessell_1988,
       author = {{Bessell}, M.~S. and {Brett}, J.~M.},
        title = "{JHKLM Photometry: Standard Systems, Passbands, and Intrinsic Colors}",
      journal = {\pasp},
         year = 1988,
        month = sep,
       volume = {100},
        pages = {1134},
          doi = {10.1086/132281},
       adsurl = {https://ui.adsabs.harvard.edu/abs/1988PASP..100.1134B}
}

@ARTICLE{Meyer_1997,
       author = {{Meyer}, Michael R. and {Calvet}, Nuria and {Hillenbrand}, Lynne A.},
        title = "{Intrinsic Near-Infrared Excesses of T Tauri Stars: Understanding the Classical T Tauri Star Locus}",
      journal = {\aj},
         year = 1997,
        month = jul,
       volume = {114},
        pages = {288-300},
          doi = {10.1086/118474},
       adsurl = {https://ui.adsabs.harvard.edu/abs/1997AJ....114..288M}
}

@ARTICLE{Anastasopoulou_2025,
       author = {{Anastasopoulou}, K. and {Guarcello}, M.~G. and {Drake}, J.~J. and {Ritchie}, B. and {De Becker}, M. and {Bayo}, A. and {Najarro}, F. and {Negueruela}, I. and {Sciortino}, S. and {Flaccomio}, E. and {Castellanos}, R. and {Albacete-Colombo}, J.~F. and {Andersen}, M. and {Damiani}, F. and {Fraschetti}, F. and {Gennaro}, M. and {Gunderson}, S.~J. and {Larkin}, C.~J.~K. and {Mackey}, J. and {Moffat}, A.~F.~J. and {Pradhan}, P. and {Saracino}, S. and {Stevens}, I.~R. and {Weigelt}, G.},
        title = "{EWOCS-IV: 1Ms ACIS Chandra observation of the supergiant B[e] star Wd1-9}",
      journal = {\aap},
         year = 2025,
        month = sep,
       volume = {701},
          eid = {A138},
        pages = {A138},
          doi = {10.1051/0004-6361/202555305},
archivePrefix = {arXiv},
       eprint = {2507.17816},
 primaryClass = {astro-ph.HE},
       adsurl = {https://ui.adsabs.harvard.edu/abs/2025A&A...701A.138A}
}

@ARTICLE{Guo_2022,
       author = {{Guo}, Zhen and {Lucas}, P.~W. and {Smith}, L.~C. and {Clarke}, C. and {Contreras Pe{\~n}a}, C. and {Bayo}, A. and {Brice{\~n}o}, C. and {Elias}, J. and {Kurtev}, R.~G. and {Borissova}, J. and {Alonso-Garc{\'\i}a}, J. and {Minniti}, D. and {Catelan}, M. and {Nikzat}, F. and {Morris}, C. and {Miller}, N.},
        title = "{Large-amplitude periodic outbursts and long-period variables in the VVV VIRAC2-{\ensuremath{\beta}} data base}",
      journal = {\mnras},
         year = 2022,
        month = jun,
       volume = {513},
       number = {1},
        pages = {1015-1035},
          doi = {10.1093/mnras/stac768},
archivePrefix = {arXiv},
       eprint = {2203.08681},
 primaryClass = {astro-ph.SR},
       adsurl = {https://ui.adsabs.harvard.edu/abs/2022MNRAS.513.1015G}
}

@ARTICLE{Gaia_2018,
       author = {{Gaia Collaboration} and {Brown}, A.~G.~A. and {Vallenari}, A. and {Prusti}, T. and {de Bruijne}, J.~H.~J. and {Babusiaux}, C. and {Bailer-Jones}, C.~A.~L. and {Biermann}, M. and {Evans}, D.~W. and {Eyer}, L. and {Jansen}, F. and {Jordi}, C. and {Klioner}, S.~A. and {Lammers}, U. and {Lindegren}, L. and {Luri}, X. and {Mignard}, F. and {Panem}, C. and {Pourbaix}, D. and {Randich}, S. and {Sartoretti}, P. and {Siddiqui}, H.~I. and {Soubiran}, C. and {van Leeuwen}, F. and {Walton}, N.~A. and {Arenou}, F. and {Bastian}, U. and {Cropper}, M. and {Drimmel}, R. and {Katz}, D. and {Lattanzi}, M.~G. and {Bakker}, J. and {Cacciari}, C. and {Casta{\~n}eda}, J. and {Chaoul}, L. and {Cheek}, N. and {De Angeli}, F. and {Fabricius}, C. and {Guerra}, R. and {Holl}, B. and {Masana}, E. and {Messineo}, R. and {Mowlavi}, N. and {Nienartowicz}, K. and {Panuzzo}, P. and {Portell}, J. and {Riello}, M. and {Seabroke}, G.~M. and {Tanga}, P. and {Th{\'e}venin}, F. and {Gracia-Abril}, G. and {Comoretto}, G. and {Garcia-Reinaldos}, M. and {Teyssier}, D. and {Altmann}, M. and {Andrae}, R. and {Audard}, M. and {Bellas-Velidis}, I. and {Benson}, K. and {Berthier}, J. and {Blomme}, R. and {Burgess}, P. and {Busso}, G. and {Carry}, B. and {Cellino}, A. and {Clementini}, G. and {Clotet}, M. and {Creevey}, O. and {Davidson}, M. and {De Ridder}, J. and {Delchambre}, L. and {Dell'Oro}, A. and {Ducourant}, C. and {Fern{\'a}ndez-Hern{\'a}ndez}, J. and {Fouesneau}, M. and {Fr{\'e}mat}, Y. and {Galluccio}, L. and {Garc{\'\i}a-Torres}, M. and {Gonz{\'a}lez-N{\'u}{\~n}ez}, J. and {Gonz{\'a}lez-Vidal}, J.~J. and {Gosset}, E. and {Guy}, L.~P. and {Halbwachs}, J. -L. and {Hambly}, N.~C. and {Harrison}, D.~L. and {Hern{\'a}ndez}, J. and {Hestroffer}, D. and {Hodgkin}, S.~T. and {Hutton}, A. and {Jasniewicz}, G. and {Jean-Antoine-Piccolo}, A. and {Jordan}, S. and {Korn}, A.~J. and {Krone-Martins}, A. and {Lanzafame}, A.~C. and {Lebzelter}, T. and {L{\"o}ffler}, W. and {Manteiga}, M. and {Marrese}, P.~M. and {Mart{\'\i}n-Fleitas}, J.~M. and {Moitinho}, A. and {Mora}, A. and {Muinonen}, K. and {Osinde}, J. and {Pancino}, E. and {Pauwels}, T. and {Petit}, J. -M. and {Recio-Blanco}, A. and {Richards}, P.~J. and {Rimoldini}, L. and {Robin}, A.~C. and {Sarro}, L.~M. and {Siopis}, C. and {Smith}, M. and {Sozzetti}, A. and {S{\"u}veges}, M. and {Torra}, J. and {van Reeven}, W. and {Abbas}, U. and {Abreu Aramburu}, A. and {Accart}, S. and {Aerts}, C. and {Altavilla}, G. and {{\'A}lvarez}, M.~A. and {Alvarez}, R. and {Alves}, J. and {Anderson}, R.~I. and {Andrei}, A.~H. and {Anglada Varela}, E. and {Antiche}, E. and {Antoja}, T. and {Arcay}, B. and {Astraatmadja}, T.~L. and {Bach}, N. and {Baker}, S.~G. and {Balaguer-N{\'u}{\~n}ez}, L. and {Balm}, P. and {Barache}, C. and {Barata}, C. and {Barbato}, D. and {Barblan}, F. and {Barklem}, P.~S. and {Barrado}, D. and {Barros}, M. and {Barstow}, M.~A. and {Bartholom{\'e} Mu{\~n}oz}, S. and {Bassilana}, J. -L. and {Becciani}, U. and {Bellazzini}, M. and {Berihuete}, A. and {Bertone}, S. and {Bianchi}, L. and {Bienaym{\'e}}, O. and {Blanco-Cuaresma}, S. and {Boch}, T. and {Boeche}, C. and {Bombrun}, A. and {Borrachero}, R. and {Bossini}, D. and {Bouquillon}, S. and {Bourda}, G. and {Bragaglia}, A. and {Bramante}, L. and {Breddels}, M.~A. and {Bressan}, A. and {Brouillet}, N. and {Br{\"u}semeister}, T. and {Brugaletta}, E. and {Bucciarelli}, B. and {Burlacu}, A. and {Busonero}, D. and {Butkevich}, A.~G. and {Buzzi}, R. and {Caffau}, E. and {Cancelliere}, R. and {Cannizzaro}, G. and {Cantat-Gaudin}, T. and {Carballo}, R. and {Carlucci}, T. and {Carrasco}, J.~M. and {Casamiquela}, L. and {Castellani}, M. and {Castro-Ginard}, A. and {Charlot}, P. and {Chemin}, L. and {Chiavassa}, A. and {Cocozza}, G. and {Costigan}, G. and {Cowell}, S. and {Crifo}, F. and {Crosta}, M. and {Crowley}, C. and {Cuypers}, J. and {Dafonte}, C. and {Damerdji}, Y. and {Dapergolas}, A. and {David}, P. and {David}, M. and {de Laverny}, P. and {De Luise}, F. and {De March}, R. and {de Martino}, D. and {de Souza}, R. and {de Torres}, A. and {Debosscher}, J. and {del Pozo}, E. and {Delbo}, M. and {Delgado}, A. and {Delgado}, H.~E. and {Di Matteo}, P. and {Diakite}, S. and {Diener}, C. and {Distefano}, E. and {Dolding}, C. and {Drazinos}, P. and {Dur{\'a}n}, J. and {Edvardsson}, B. and {Enke}, H. and {Eriksson}, K. and {Esquej}, P. and {Eynard Bontemps}, G. and {Fabre}, C. and {Fabrizio}, M. and {Faigler}, S. and {Falc{\~a}o}, A.~J. and {Farr{\`a}s Casas}, M. and {Federici}, L. and {Fedorets}, G. and {Fernique}, P. and {Figueras}, F. and {Filippi}, F. and {Findeisen}, K. and {Fonti}, A. and {Fraile}, E. and {Fraser}, M. and {Fr{\'e}zouls}, B. and {Gai}, M. and {Galleti}, S. and {Garabato}, D. and {Garc{\'\i}a-Sedano}, F. and {Garofalo}, A. and {Garralda}, N. and {Gavel}, A. and {Gavras}, P. and {Gerssen}, J. and {Geyer}, R. and {Giacobbe}, P. and {Gilmore}, G. and {Girona}, S. and {Giuffrida}, G. and {Glass}, F. and {Gomes}, M. and {Granvik}, M. and {Gueguen}, A. and {Guerrier}, A. and {Guiraud}, J. and {Guti{\'e}rrez-S{\'a}nchez}, R. and {Haigron}, R. and {Hatzidimitriou}, D. and {Hauser}, M. and {Haywood}, M. and {Heiter}, U. and {Helmi}, A. and {Heu}, J. and {Hilger}, T. and {Hobbs}, D. and {Hofmann}, W. and {Holland}, G. and {Huckle}, H.~E. and {Hypki}, A. and {Icardi}, V. and {Jan{\ss}en}, K. and {Jevardat de Fombelle}, G. and {Jonker}, P.~G. and {Juh{\'a}sz}, {\'A}. L. and {Julbe}, F. and {Karampelas}, A. and {Kewley}, A. and {Klar}, J. and {Kochoska}, A. and {Kohley}, R. and {Kolenberg}, K. and {Kontizas}, M. and {Kontizas}, E. and {Koposov}, S.~E. and {Kordopatis}, G. and {Kostrzewa-Rutkowska}, Z. and {Koubsky}, P. and {Lambert}, S. and {Lanza}, A.~F. and {Lasne}, Y. and {Lavigne}, J. -B. and {Le Fustec}, Y. and {Le Poncin-Lafitte}, C. and {Lebreton}, Y. and {Leccia}, S. and {Leclerc}, N. and {Lecoeur-Taibi}, I. and {Lenhardt}, H. and {Leroux}, F. and {Liao}, S. and {Licata}, E. and {Lindstr{\o}m}, H.~E.~P. and {Lister}, T.~A. and {Livanou}, E. and {Lobel}, A. and {L{\'o}pez}, M. and {Managau}, S. and {Mann}, R.~G. and {Mantelet}, G. and {Marchal}, O. and {Marchant}, J.~M. and {Marconi}, M. and {Marinoni}, S. and {Marschalk{\'o}}, G. and {Marshall}, D.~J. and {Martino}, M. and {Marton}, G. and {Mary}, N. and {Massari}, D. and {Matijevi{\v{c}}}, G. and {Mazeh}, T. and {McMillan}, P.~J. and {Messina}, S. and {Michalik}, D. and {Millar}, N.~R. and {Molina}, D. and {Molinaro}, R. and {Moln{\'a}r}, L. and {Montegriffo}, P. and {Mor}, R. and {Morbidelli}, R. and {Morel}, T. and {Morris}, D. and {Mulone}, A.~F. and {Muraveva}, T. and {Musella}, I. and {Nelemans}, G. and {Nicastro}, L. and {Noval}, L. and {O'Mullane}, W. and {Ord{\'e}novic}, C. and {Ord{\'o}{\~n}ez-Blanco}, D. and {Osborne}, P. and {Pagani}, C. and {Pagano}, I. and {Pailler}, F. and {Palacin}, H. and {Palaversa}, L. and {Panahi}, A. and {Pawlak}, M. and {Piersimoni}, A.~M. and {Pineau}, F. -X. and {Plachy}, E. and {Plum}, G. and {Poggio}, E. and {Poujoulet}, E. and {Pr{\v{s}}a}, A. and {Pulone}, L. and {Racero}, E. and {Ragaini}, S. and {Rambaux}, N. and {Ramos-Lerate}, M. and {Regibo}, S. and {Reyl{\'e}}, C. and {Riclet}, F. and {Ripepi}, V. and {Riva}, A. and {Rivard}, A. and {Rixon}, G. and {Roegiers}, T. and {Roelens}, M. and {Romero-G{\'o}mez}, M. and {Rowell}, N. and {Royer}, F. and {Ruiz-Dern}, L. and {Sadowski}, G. and {Sagrist{\`a} Sell{\'e}s}, T. and {Sahlmann}, J. and {Salgado}, J. and {Salguero}, E. and {Sanna}, N. and {Santana-Ros}, T. and {Sarasso}, M. and {Savietto}, H. and {Schultheis}, M. and {Sciacca}, E. and {Segol}, M. and {Segovia}, J.~C. and {S{\'e}gransan}, D. and {Shih}, I. -C. and {Siltala}, L. and {Silva}, A.~F. and {Smart}, R.~L. and {Smith}, K.~W. and {Solano}, E. and {Solitro}, F. and {Sordo}, R. and {Soria Nieto}, S. and {Souchay}, J. and {Spagna}, A. and {Spoto}, F. and {Stampa}, U. and {Steele}, I.~A. and {Steidelm{\"u}ller}, H. and {Stephenson}, C.~A. and {Stoev}, H. and {Suess}, F.~F. and {Surdej}, J. and {Szabados}, L. and {Szegedi-Elek}, E. and {Tapiador}, D. and {Taris}, F. and {Tauran}, G. and {Taylor}, M.~B. and {Teixeira}, R. and {Terrett}, D. and {Teyssandier}, P. and {Thuillot}, W. and {Titarenko}, A. and {Torra Clotet}, F. and {Turon}, C. and {Ulla}, A. and {Utrilla}, E. and {Uzzi}, S. and {Vaillant}, M. and {Valentini}, G. and {Valette}, V. and {van Elteren}, A. and {Van Hemelryck}, E. and {van Leeuwen}, M. and {Vaschetto}, M. and {Vecchiato}, A. and {Veljanoski}, J. and {Viala}, Y. and {Vicente}, D. and {Vogt}, S. and {von Essen}, C. and {Voss}, H. and {Votruba}, V. and {Voutsinas}, S. and {Walmsley}, G. and {Weiler}, M. and {Wertz}, O. and {Wevers}, T. and {Wyrzykowski}, {\L}. and {Yoldas}, A. and {{\v{Z}}erjal}, M. and {Ziaeepour}, H. and {Zorec}, J. and {Zschocke}, S. and {Zucker}, S. and {Zurbach}, C. and {Zwitter}, T.},
        title = "{Gaia Data Release 2. Summary of the contents and survey properties}",
      journal = {\aap},
         year = 2018,
        month = aug,
       volume = {616},
          eid = {A1},
        pages = {A1},
          doi = {10.1051/0004-6361/201833051},
archivePrefix = {arXiv},
       eprint = {1804.09365},
 primaryClass = {astro-ph.GA},
       adsurl = {https://ui.adsabs.harvard.edu/abs/2018A&A...616A...1G}
}

@ARTICLE{Clark_2010,
       author = {{Clark}, J.~S. and {Ritchie}, B.~W. and {Negueruela}, I.},
        title = "{A serendipitous survey for variability amongst the massive stellar population of Westerlund 1}",
      journal = {\aap},
         year = 2010,
        month = may,
       volume = {514},
          eid = {A87},
        pages = {A87},
          doi = {10.1051/0004-6361/200913820},
archivePrefix = {arXiv},
       eprint = {1003.5107},
 primaryClass = {astro-ph.SR},
       adsurl = {https://ui.adsabs.harvard.edu/abs/2010A&A...514A..87C}
}

@ARTICLE{Cantat-Gaudin_2020,
       author = {{Cantat-Gaudin}, T. and {Anders}, F.},
        title = "{Clusters and mirages: cataloguing stellar aggregates in the Milky Way}",
      journal = {\aap},
         year = 2020,
        month = jan,
       volume = {633},
          eid = {A99},
        pages = {A99},
          doi = {10.1051/0004-6361/201936691},
archivePrefix = {arXiv},
       eprint = {1911.07075},
 primaryClass = {astro-ph.SR},
       adsurl = {https://ui.adsabs.harvard.edu/abs/2020A&A...633A..99C}
}

@ARTICLE{Lomb1976,
       author = {{Lomb}, N.~R.},
        title = "{Least-Squares Frequency Analysis of Unequally Spaced Data}",
      journal = {\apss},
         year = 1976,
        month = feb,
       volume = {39},
       number = {2},
        pages = {447-462},
          doi = {10.1007/BF00648343},
       adsurl = {https://ui.adsabs.harvard.edu/abs/1976Ap&SS..39..447L}
}

@ARTICLE{Scargle1982,
       author = {{Scargle}, J.~D.},
        title = "{Studies in astronomical time series analysis. II. Statistical aspects of spectral analysis of unevenly spaced data.}",
      journal = {\apj},
         year = 1982,
        month = dec,
       volume = {263},
        pages = {835-853},
          doi = {10.1086/160554},
       adsurl = {https://ui.adsabs.harvard.edu/abs/1982ApJ...263..835S}
}

@ARTICLE{Gennaro_2011,
       author = {{Gennaro}, M. and {Brandner}, W. and {Stolte}, A. and {Henning}, Th.},
        title = "{Mass segregation and elongation of the starburst cluster Westerlund 1}",
      journal = {\mnras},
         year = 2011,
        month = apr,
       volume = {412},
       number = {4},
        pages = {2469-2488},
          doi = {10.1111/j.1365-2966.2010.18068.x},
archivePrefix = {arXiv},
       eprint = {1011.5223},
 primaryClass = {astro-ph.GA},
       adsurl = {https://ui.adsabs.harvard.edu/abs/2011MNRAS.412.2469G}
}

@ARTICLE{Lim_2013,
       author = {{Lim}, Beomdu and {Chun}, Moo-Young and {Sung}, Hwankyung and {Park}, Byeong-Gon and {Lee}, Jae-Joon and {Sohn}, Sangmo T. and {Hur}, Hyeonoh and {Bessell}, Michael S.},
        title = "{The Starburst Cluster Westerlund 1: The Initial Mass Function and Mass Segregation}",
      journal = {\aj},
         year = 2013,
        month = feb,
       volume = {145},
       number = {2},
          eid = {46},
        pages = {46},
          doi = {10.1088/0004-6256/145/2/46},
archivePrefix = {arXiv},
       eprint = {1211.5832},
 primaryClass = {astro-ph.SR},
       adsurl = {https://ui.adsabs.harvard.edu/abs/2013AJ....145...46L}
}

@ARTICLE{Molnar_2022,
       author = {{Molnar}, Thomas A. and {Sanders}, Jason L. and {Smith}, Leigh C. and {Belokurov}, Vasily and {Lucas}, Philip and {Minniti}, Dante},
        title = "{Variable star classification across the Galactic bulge and disc with the VISTA Variables in the V{\'\i}a L{\'a}ctea survey}",
      journal = {\mnras},
         year = 2022,
        month = jan,
       volume = {509},
       number = {2},
        pages = {2566-2592},
          doi = {10.1093/mnras/stab3116},
archivePrefix = {arXiv},
       eprint = {2110.15371},
 primaryClass = {astro-ph.SR},
       adsurl = {https://ui.adsabs.harvard.edu/abs/2022MNRAS.509.2566M}
}

@ARTICLE{Ritchie_2022,
       author = {{Ritchie}, B.~W. and {Clark}, J.~S. and {Negueruela}, I. and {Najarro}, F.},
        title = "{A VLT/FLAMES survey for massive binaries in Westerlund 1. VIII. Binary systems and orbital parameters}",
      journal = {\aap},
         year = 2022,
        month = apr,
       volume = {660},
          eid = {A89},
        pages = {A89},
          doi = {10.1051/0004-6361/202142405},
archivePrefix = {arXiv},
       eprint = {2111.12463},
 primaryClass = {astro-ph.SR},
       adsurl = {https://ui.adsabs.harvard.edu/abs/2022A&A...660A..89R}
}

@ARTICLE{Cody_2014,
       author = {{Cody}, Ann Marie and {Stauffer}, John and {Baglin}, Annie and {Micela}, Giuseppina and {Rebull}, Luisa M. and {Flaccomio}, Ettore and {Morales-Calder{\'o}n}, Mar{\'\i}a and {Aigrain}, Suzanne and {Bouvier}, J{\`e}r{\^o}me and {Hillenbrand}, Lynne A. and {Gutermuth}, Robert and {Song}, Inseok and {Turner}, Neal and {Alencar}, Silvia H.~P. and {Zwintz}, Konstanze and {Plavchan}, Peter and {Carpenter}, John and {Findeisen}, Krzysztof and {Carey}, Sean and {Terebey}, Susan and {Hartmann}, Lee and {Calvet}, Nuria and {Teixeira}, Paula and {Vrba}, Frederick J. and {Wolk}, Scott and {Covey}, Kevin and {Poppenhaeger}, Katja and {G{\"u}nther}, Hans Moritz and {Forbrich}, Jan and {Whitney}, Barbara and {Affer}, Laura and {Herbst}, William and {Hora}, Joseph and {Barrado}, David and {Holtzman}, Jon and {Marchis}, Franck and {Wood}, Kenneth and {Medeiros Guimar{\~a}es}, Marcelo and {Lillo Box}, Jorge and {Gillen}, Ed and {McQuillan}, Amy and {Espaillat}, Catherine and {Allen}, Lori and {D'Alessio}, Paola and {Favata}, Fabio},
        title = "{CSI 2264: Simultaneous Optical and Infrared Light Curves of Young Disk-bearing Stars in NGC 2264 with CoRoT and Spitzer{\textemdash}Evidence for Multiple Origins of Variability}",
      journal = {\aj},
         year = 2014,
        month = apr,
       volume = {147},
       number = {4},
          eid = {82},
        pages = {82},
          doi = {10.1088/0004-6256/147/4/82},
archivePrefix = {arXiv},
       eprint = {1401.6582},
 primaryClass = {astro-ph.SR},
       adsurl = {https://ui.adsabs.harvard.edu/abs/2014AJ....147...82C}
}

@ARTICLE{Sanchez_2020,
       author = {{S{\'a}nchez}, N{\'e}stor and {Alfaro}, Emilio J. and {L{\'o}pez-Mart{\'\i}nez}, F{\'a}tima},
        title = "{A catalogue of open cluster radii determined from Gaia proper motions}",
      journal = {\mnras},
         year = 2020,
        month = jul,
       volume = {495},
       number = {3},
        pages = {2882-2893},
          doi = {10.1093/mnras/staa1359},
archivePrefix = {arXiv},
       eprint = {2005.05924},
 primaryClass = {astro-ph.GA},
       adsurl = {https://ui.adsabs.harvard.edu/abs/2020MNRAS.495.2882S}
}

@ARTICLE{Yuk_2022,
       author = {{Yuk}, Heechan and {Dai}, Xinyu and {Jayasinghe}, T. and {Fu}, Hai and {Mishra}, Hora D. and {Kochanek}, Christopher S. and {Shappee}, Benjamin J. and {Stanek}, K.~Z.},
        title = "{Variability Selected Active Galactic Nuclei from ASAS-SN Survey: Constraining the Low Luminosity AGN Population}",
      journal = {\apj},
         year = 2022,
        month = may,
       volume = {930},
       number = {2},
          eid = {110},
        pages = {110},
          doi = {10.3847/1538-4357/ac6423},
archivePrefix = {arXiv},
       eprint = {2203.08348},
 primaryClass = {astro-ph.GA},
       adsurl = {https://ui.adsabs.harvard.edu/abs/2022ApJ...930..110Y}
}

@ARTICLE{Dias_2002,
       author = {{Dias}, W.~S. and {Alessi}, B.~S. and {Moitinho}, A. and {L{\'e}pine}, J.~R.~D.},
        title = "{New catalogue of optically visible open clusters and candidates}",
      journal = {\aap},
         year = 2002,
        month = jul,
       volume = {389},
        pages = {871-873},
          doi = {10.1051/0004-6361:20020668},
archivePrefix = {arXiv},
       eprint = {astro-ph/0203351},
 primaryClass = {astro-ph},
       adsurl = {https://ui.adsabs.harvard.edu/abs/2002A&A...389..871D}
}

@ARTICLE{Guarcello_2024,
       author = {{Guarcello}, M.~G. and {Flaccomio}, E. and {Albacete-Colombo}, J.~F. and {Almendros-Abad}, V. and {Anastasopoulou}, K. and {Andersen}, M. and {Argiroffi}, C. and {Bayo}, A. and {Bartlett}, E.~S. and {Bastian}, N. and {De Becker}, M. and {Best}, W. and {Bonito}, R. and {Borghese}, A. and {Calzetti}, D. and {Castellanos}, R. and {Cecchi-Pestellini}, C. and {Clark}, J.~S. and {Clarke}, C.~J. and {Coti Zelati}, F. and {Damiani}, F. and {Drake}, J.~J. and {Gennaro}, M. and {Ginsburg}, A. and {Grebel}, E.~K. and {Hora}, J.~L. and {Israel}, G.~L. and {Lawrence}, G. and {Locci}, D. and {Mapelli}, M. and {Martinez-Galarza}, J.~R. and {Micela}, G. and {Miceli}, M. and {Moraux}, E. and {Muzic}, K. and {Najarro}, F. and {Negueruela}, I. and {Nota}, A. and {Pallanca}, C. and {Prisinzano}, L. and {Ritchie}, B. and {Robberto}, M. and {Rom}, T. and {Sabbi}, E. and {Scholz}, A. and {Sciortino}, S. and {Trigilio}, C. and {Umana}, G. and {Winter}, A. and {Wright}, N.~J. and {Zeidler}, P.},
        title = "{EWOCS-I: The catalog of X-ray sources in Westerlund 1 from the Extended Westerlund 1 and 2 Open Clusters Survey}",
      journal = {\aap},
         year = 2024,
        month = feb,
       volume = {682},
          eid = {A49},
        pages = {A49},
          doi = {10.1051/0004-6361/202347778},
archivePrefix = {arXiv},
       eprint = {2312.08947},
 primaryClass = {astro-ph.SR},
       adsurl = {https://ui.adsabs.harvard.edu/abs/2024A&A...682A..49G}
}

@ARTICLE{Guarcello_2025,
       author = {{Guarcello}, M.~G. and {Almendros-Abad}, V. and {Lovell}, J.~B. and {Monsch}, K. and {Mu{\v{z}}i{\'c}}, K. and {Mart{\'\i}nez-Galarza}, J.~R. and {Drake}, J.~J. and {Anastasopoulou}, K. and {Andersen}, M. and {Argiroffi}, C. and {Bayo}, A. and {Bonito}, R. and {Capela}, D. and {Damiani}, F. and {Gennaro}, M. and {Ginsburg}, A. and {Grebel}, E.~K. and {Hora}, J.~L. and {Moraux}, E. and {Najarro}, F. and {Negueruela}, I. and {Prisinzano}, L. and {Richardson}, N.~D. and {Ritchie}, B. and {Robberto}, M. and {Rom}, T. and {Sabbi}, E. and {Sciortino}, S. and {Umana}, G. and {Winter}, A. and {Wright}, N.~J. and {Zeidler}, P.},
        title = "{EWOCS-III: JWST observations of the supermassive star cluster Westerlund 1}",
      journal = {\aap},
         year = 2025,
        month = jan,
       volume = {693},
          eid = {A120},
        pages = {A120},
          doi = {10.1051/0004-6361/202452150},
archivePrefix = {arXiv},
       eprint = {2411.13051},
 primaryClass = {astro-ph.SR},
       adsurl = {https://ui.adsabs.harvard.edu/abs/2025A&A...693A.120G}
}

@ARTICLE{Sanchez_2010,
       author = {{S{\'a}nchez}, N. and {Vicente}, B. and {Alfaro}, E.~J.},
        title = "{Cluster radius and sampling radius in the determination of cluster membership probabilities}",
      journal = {\aap},
         year = 2010,
        month = feb,
       volume = {510},
          eid = {A78},
        pages = {A78},
          doi = {10.1051/0004-6361/200912886},
archivePrefix = {arXiv},
       eprint = {0911.5693},
 primaryClass = {astro-ph.GA},
       adsurl = {https://ui.adsabs.harvard.edu/abs/2010A&A...510A..78S}
}

@ARTICLE{Bonanos_2007,
       author = {{Bonanos}, Alceste Z.},
        title = "{Variability of Young Massive Stars in the Galactic Super Star Cluster Westerlund 1}",
      journal = {\aj},
         year = 2007,
        month = jun,
       volume = {133},
       number = {6},
        pages = {2696-2708},
          doi = {10.1086/518093},
archivePrefix = {arXiv},
       eprint = {astro-ph/0702614},
 primaryClass = {astro-ph},
       adsurl = {https://ui.adsabs.harvard.edu/abs/2007AJ....133.2696B}
}

@ARTICLE{Negueruela_2010,
       author = {{Negueruela}, I. and {Clark}, J.~S. and {Ritchie}, B.~W.},
        title = "{The population of OB supergiants in the starburst cluster Westerlund 1}",
      journal = {\aap},
         year = 2010,
        month = jun,
       volume = {516},
          eid = {A78},
        pages = {A78},
          doi = {10.1051/0004-6361/201014032},
archivePrefix = {arXiv},
       eprint = {1003.5204},
 primaryClass = {astro-ph.GA},
       adsurl = {https://ui.adsabs.harvard.edu/abs/2010A&A...516A..78N}
}

@ARTICLE{Gennaro_2017,
       author = {{Gennaro}, Mario and {Goodwin}, Simon P. and {Parker}, Richard J. and {Allison}, Richard J. and {Brandner}, Wolfgang},
        title = "{Hierarchical formation of Westerlund 1: a collapsing cluster with no primordial mass segregation?}",
      journal = {\mnras},
         year = 2017,
        month = dec,
       volume = {472},
       number = {2},
        pages = {1760-1769},
          doi = {10.1093/mnras/stx2098},
archivePrefix = {arXiv},
       eprint = {1708.04161},
 primaryClass = {astro-ph.GA},
       adsurl = {https://ui.adsabs.harvard.edu/abs/2017MNRAS.472.1760G}
}

@ARTICLE{Gaia_2023,
       author = {{Gaia Collaboration} and {Vallenari}, A. and {Brown}, A.~G.~A. and {Prusti}, T. and {de Bruijne}, J.~H.~J. and {Arenou}, F. and {Babusiaux}, C. and {Biermann}, M. and {Creevey}, O.~L. and {Ducourant}, C. and {Evans}, D.~W. and {Eyer}, L. and {Guerra}, R. and {Hutton}, A. and {Jordi}, C. and {Klioner}, S.~A. and {Lammers}, U.~L. and {Lindegren}, L. and {Luri}, X. and {Mignard}, F. and {Panem}, C. and {Pourbaix}, D. and {Randich}, S. and {Sartoretti}, P. and {Soubiran}, C. and {Tanga}, P. and {Walton}, N.~A. and {Bailer-Jones}, C.~A.~L. and {Bastian}, U. and {Drimmel}, R. and {Jansen}, F. and {Katz}, D. and {Lattanzi}, M.~G. and {van Leeuwen}, F. and {Bakker}, J. and {Cacciari}, C. and {Casta{\~n}eda}, J. and {De Angeli}, F. and {Fabricius}, C. and {Fouesneau}, M. and {Fr{\'e}mat}, Y. and {Galluccio}, L. and {Guerrier}, A. and {Heiter}, U. and {Masana}, E. and {Messineo}, R. and {Mowlavi}, N. and {Nicolas}, C. and {Nienartowicz}, K. and {Pailler}, F. and {Panuzzo}, P. and {Riclet}, F. and {Roux}, W. and {Seabroke}, G.~M. and {Sordo}, R. and {Th{\'e}venin}, F. and {Gracia-Abril}, G. and {Portell}, J. and {Teyssier}, D. and {Altmann}, M. and {Andrae}, R. and {Audard}, M. and {Bellas-Velidis}, I. and {Benson}, K. and {Berthier}, J. and {Blomme}, R. and {Burgess}, P.~W. and {Busonero}, D. and {Busso}, G. and {C{\'a}novas}, H. and {Carry}, B. and {Cellino}, A. and {Cheek}, N. and {Clementini}, G. and {Damerdji}, Y. and {Davidson}, M. and {de Teodoro}, P. and {Nu{\~n}ez Campos}, M. and {Delchambre}, L. and {Dell'Oro}, A. and {Esquej}, P. and {Fern{\'a}ndez-Hern{\'a}ndez}, J. and {Fraile}, E. and {Garabato}, D. and {Garc{\'\i}a-Lario}, P. and {Gosset}, E. and {Haigron}, R. and {Halbwachs}, J. -L. and {Hambly}, N.~C. and {Harrison}, D.~L. and {Hern{\'a}ndez}, J. and {Hestroffer}, D. and {Hodgkin}, S.~T. and {Holl}, B. and {Jan{\ss}en}, K. and {Jevardat de Fombelle}, G. and {Jordan}, S. and {Krone-Martins}, A. and {Lanzafame}, A.~C. and {L{\"o}ffler}, W. and {Marchal}, O. and {Marrese}, P.~M. and {Moitinho}, A. and {Muinonen}, K. and {Osborne}, P. and {Pancino}, E. and {Pauwels}, T. and {Recio-Blanco}, A. and {Reyl{\'e}}, C. and {Riello}, M. and {Rimoldini}, L. and {Roegiers}, T. and {Rybizki}, J. and {Sarro}, L.~M. and {Siopis}, C. and {Smith}, M. and {Sozzetti}, A. and {Utrilla}, E. and {van Leeuwen}, M. and {Abbas}, U. and {{\'A}brah{\'a}m}, P. and {Abreu Aramburu}, A. and {Aerts}, C. and {Aguado}, J.~J. and {Ajaj}, M. and {Aldea-Montero}, F. and {Altavilla}, G. and {{\'A}lvarez}, M.~A. and {Alves}, J. and {Anders}, F. and {Anderson}, R.~I. and {Anglada Varela}, E. and {Antoja}, T. and {Baines}, D. and {Baker}, S.~G. and {Balaguer-N{\'u}{\~n}ez}, L. and {Balbinot}, E. and {Balog}, Z. and {Barache}, C. and {Barbato}, D. and {Barros}, M. and {Barstow}, M.~A. and {Bartolom{\'e}}, S. and {Bassilana}, J. -L. and {Bauchet}, N. and {Becciani}, U. and {Bellazzini}, M. and {Berihuete}, A. and {Bernet}, M. and {Bertone}, S. and {Bianchi}, L. and {Binnenfeld}, A. and {Blanco-Cuaresma}, S. and {Blazere}, A. and {Boch}, T. and {Bombrun}, A. and {Bossini}, D. and {Bouquillon}, S. and {Bragaglia}, A. and {Bramante}, L. and {Breedt}, E. and {Bressan}, A. and {Brouillet}, N. and {Brugaletta}, E. and {Bucciarelli}, B. and {Burlacu}, A. and {Butkevich}, A.~G. and {Buzzi}, R. and {Caffau}, E. and {Cancelliere}, R. and {Cantat-Gaudin}, T. and {Carballo}, R. and {Carlucci}, T. and {Carnerero}, M.~I. and {Carrasco}, J.~M. and {Casamiquela}, L. and {Castellani}, M. and {Castro-Ginard}, A. and {Chaoul}, L. and {Charlot}, P. and {Chemin}, L. and {Chiaramida}, V. and {Chiavassa}, A. and {Chornay}, N. and {Comoretto}, G. and {Contursi}, G. and {Cooper}, W.~J. and {Cornez}, T. and {Cowell}, S. and {Crifo}, F. and {Cropper}, M. and {Crosta}, M. and {Crowley}, C. and {Dafonte}, C. and {Dapergolas}, A. and {David}, M. and {David}, P. and {de Laverny}, P. and {De Luise}, F. and {De March}, R.},
        title = "{Gaia Data Release 3. Summary of the content and survey properties}",
      journal = {\aap},
         year = 2023,
        month = jun,
       volume = {674},
          eid = {A1},
        pages = {A1},
          doi = {10.1051/0004-6361/202243940},
archivePrefix = {arXiv},
       eprint = {2208.00211},
 primaryClass = {astro-ph.GA},
       adsurl = {https://ui.adsabs.harvard.edu/abs/2023A&A...674A...1G}
}

@ARTICLE{Marsh_2017,
       author = {{Marsh}, K.~A. and {Whitworth}, A.~P. and {Lomax}, O. and {Ragan}, S.~E. and {Becciani}, U. and {Cambr{\'e}sy}, L. and {Di Giorgio}, A. and {Eden}, D. and {Elia}, D. and {Kacsuk}, P. and {Molinari}, S. and {Palmeirim}, P. and {Pezzuto}, S. and {Schneider}, N. and {Sciacca}, E. and {Vitello}, F.},
        title = "{Multitemperature mapping of dust structures throughout the Galactic Plane using the PPMAP tool with Herschel Hi-GAL data}",
      journal = {\mnras},
         year = 2017,
        month = nov,
       volume = {471},
       number = {3},
        pages = {2730-2742},
          doi = {10.1093/mnras/stx1723},
archivePrefix = {arXiv},
       eprint = {1707.03808},
 primaryClass = {astro-ph.GA},
       adsurl = {https://ui.adsabs.harvard.edu/abs/2017MNRAS.471.2730M}
}

@ARTICLE{Koumpia_2012,
       author = {{Koumpia}, E. and {Bonanos}, A.~Z.},
        title = "{Fundamental parameters of four massive eclipsing binaries in Westerlund 1}",
      journal = {\aap},
         year = 2012,
        month = nov,
       volume = {547},
          eid = {A30},
        pages = {A30},
          doi = {10.1051/0004-6361/201219465},
archivePrefix = {arXiv},
       eprint = {1108.4453},
 primaryClass = {astro-ph.SR},
       adsurl = {https://ui.adsabs.harvard.edu/abs/2012A&A...547A..30K}
}

@ARTICLE{Anderson_2025,
       author = {{Anderson}, Richard I. and {Hunt}, Emily L.},
        title = "{A bird's eye view of stellar evolution through populations of variable stars in Galactic open clusters}",
      journal = {\aap},
         year = 2025,
        month = aug,
       volume = {700},
          eid = {L13},
        pages = {L13},
          doi = {10.1051/0004-6361/202555111},
archivePrefix = {arXiv},
       eprint = {2508.12866},
 primaryClass = {astro-ph.SR},
       adsurl = {https://ui.adsabs.harvard.edu/abs/2025A&A...700L..13A}
}

@ARTICLE{Clark_2020,
       author = {{Clark}, J.~S. and {Ritchie}, B.~W. and {Negueruela}, I.},
        title = "{A VLT/FLAMES survey for massive binaries in Westerlund 1. VII. Cluster census}",
      journal = {\aap},
         year = 2020,
        month = mar,
       volume = {635},
          eid = {A187},
        pages = {A187},
          doi = {10.1051/0004-6361/201935903},
archivePrefix = {arXiv},
       eprint = {1908.05616},
 primaryClass = {astro-ph.SR},
       adsurl = {https://ui.adsabs.harvard.edu/abs/2020A&A...635A.187C}
}

@ARTICLE{Navarete_2022,
       author = {{Navarete}, Felipe and {Damineli}, Augusto and {Ramirez}, Aura E. and {Rocha}, Danilo F. and {Almeida}, Leonardo A.},
        title = "{Distance and age of the massive stellar cluster Westerlund 1. I. Parallax method using Gaia-EDR3}",
      journal = {\mnras},
         year = 2022,
        month = oct,
       volume = {516},
       number = {1},
        pages = {1289-1301},
          doi = {10.1093/mnras/stac2374},
archivePrefix = {arXiv},
       eprint = {2204.09414},
 primaryClass = {astro-ph.SR},
       adsurl = {https://ui.adsabs.harvard.edu/abs/2022MNRAS.516.1289N}
}

@ARTICLE{OrdenesHuanca_2022,
       author = {{Ordenes-Huanca}, C. and {Zoccali}, M. and {Bayo}, A. and {Cuadra}, J. and {Contreras Ramos}, R. and {Hillenbrand}, L.~A. and {Lacerna}, I. and {Abarzua}, S. and {Avenda{\~n}o}, C. and {Diaz}, P. and {Fernandez}, I. and {Lara}, G.},
        title = "{Infrared variability of young solar analogues in the Lagoon Nebula}",
      journal = {\mnras},
         year = 2022,
        month = dec,
       volume = {517},
       number = {4},
        pages = {6191-6204},
          doi = {10.1093/mnras/stac3049},
archivePrefix = {arXiv},
       eprint = {2210.09242},
 primaryClass = {astro-ph.SR},
       adsurl = {https://ui.adsabs.harvard.edu/abs/2022MNRAS.517.6191O}
}

@ARTICLE{Carpenter_2001,
       author = {{Carpenter}, John M. and {Hillenbrand}, Lynne A. and {Skrutskie}, M.~F.},
        title = "{Near-Infrared Photometric Variability of Stars toward the Orion A Molecular Cloud}",
      journal = {\aj},
         year = 2001,
        month = jun,
       volume = {121},
       number = {6},
        pages = {3160-3190},
          doi = {10.1086/321086},
archivePrefix = {arXiv},
       eprint = {astro-ph/0102446},
 primaryClass = {astro-ph},
       adsurl = {https://ui.adsabs.harvard.edu/abs/2001AJ....121.3160C}
}

@ARTICLE{Andersen_2017,
       author = {{Andersen}, M. and {Gennaro}, M. and {Brandner}, W. and {Stolte}, A. and {de Marchi}, G. and {Meyer}, M.~R. and {Zinnecker}, H.},
        title = "{Very low-mass stellar content of the young supermassive Galactic star cluster Westerlund 1}",
      journal = {\aap},
         year = 2017,
        month = jun,
       volume = {602},
          eid = {A22},
        pages = {A22},
          doi = {10.1051/0004-6361/201322863},
archivePrefix = {arXiv},
       eprint = {1602.05918},
 primaryClass = {astro-ph.GA},
       adsurl = {https://ui.adsabs.harvard.edu/abs/2017A&A...602A..22A}
}

@ARTICLE{Clark_2005,
       author = {{Clark}, J.~S. and {Negueruela}, I. and {Crowther}, P.~A. and {Goodwin}, S.~P.},
        title = "{On the massive stellar population of the super star cluster <ASTROBJ>Westerlund 1</ASTROBJ>}",
      journal = {\aap},
         year = 2005,
        month = may,
       volume = {434},
       number = {3},
        pages = {949-969},
          doi = {10.1051/0004-6361:20042413},
archivePrefix = {arXiv},
       eprint = {astro-ph/0504342},
 primaryClass = {astro-ph},
       adsurl = {https://ui.adsabs.harvard.edu/abs/2005A&A...434..949C}
}

@ARTICLE{Negueruela_2022,
       author = {{Negueruela}, I. and {Alfaro}, E.~J. and {Dorda}, R. and {Marco}, A. and {Ma{\'\i}z Apell{\'a}niz}, J. and {Gonz{\'a}lez-Fern{\'a}ndez}, C.},
        title = "{Westerlund 1 under the light of Gaia EDR3: Distance, isolation, extent, and a hidden population}",
      journal = {\aap},
         year = 2022,
        month = aug,
       volume = {664},
          eid = {A146},
        pages = {A146},
          doi = {10.1051/0004-6361/202142985},
archivePrefix = {arXiv},
       eprint = {2204.00422},
 primaryClass = {astro-ph.SR},
       adsurl = {https://ui.adsabs.harvard.edu/abs/2022A&A...664A.146N}
}

@ARTICLE{Saito_2024,
       author = {{Saito}, R.~K. and {Hempel}, M. and {Alonso-Garc{\'\i}a}, J. and {Lucas}, P.~W. and {Minniti}, D. and {Alonso}, S. and {Baravalle}, L. and {Borissova}, J. and {Caceres}, C. and {Chen{\'e}}, A.~N. and {Cross}, N.~J.~G. and {Duplancic}, F. and {Garro}, E.~R. and {G{\'o}mez}, M. and {Ivanov}, V.~D. and {Kurtev}, R. and {Luna}, A. and {Majaess}, D. and {Navarro}, M.~G. and {Pullen}, J.~B. and {Rejkuba}, M. and {Sanders}, J.~L. and {Smith}, L.~C. and {Albino}, P.~H.~C. and {Alonso}, M.~V. and {Am{\^o}res}, E.~B. and {Angeloni}, R. and {Arias}, J.~I. and {Arnaboldi}, M. and {Barbuy}, B. and {Bayo}, A. and {Beamin}, J.~C. and {Bedin}, L.~R. and {Bellini}, A. and {Benjamin}, R.~A. and {Bica}, E. and {Bonatto}, C.~J. and {Botan}, E. and {Braga}, V.~F. and {Brown}, D.~A. and {Cabral}, J.~B. and {Camargo}, D. and {Caratti o Garatti}, A. and {Carballo-Bello}, J.~A. and {Catelan}, M. and {Chavero}, C. and {Chijani}, M.~A. and {Clari{\'a}}, J.~J. and {Coldwell}, G.~V. and {Contreras Pe{\~n}a}, C. and {Contreras Ramos}, R. and {Corral-Santana}, J.~M. and {Cort{\'e}s}, C.~C. and {Cort{\'e}s-Contreras}, M. and {Cruz}, P. and {Daza-Perilla}, I.~V. and {Debattista}, V.~P. and {Dias}, B. and {Donoso}, L. and {D'Souza}, R. and {Emerson}, J.~P. and {Federle}, S. and {Fermiano}, V. and {Fernandez}, J. and {Fern{\'a}ndez-Trincado}, J.~G. and {Ferreira}, T. and {Ferreira Lopes}, C.~E. and {Firpo}, V. and {Flores-Quintana}, C. and {Fraga}, L. and {Froebrich}, D. and {Galdeano}, D. and {Gavignaud}, I. and {Geisler}, D. and {Gerhard}, O.~E. and {Gieren}, W. and {Gonzalez}, O.~A. and {Gramajo}, L.~V. and {Gran}, F. and {Granitto}, P.~M. and {Griggio}, M. and {Guo}, Z. and {Gurovich}, S. and {Hilker}, M. and {Jones}, H.~R.~A. and {Kammers}, R. and {Kuhn}, M.~A. and {Kumar}, M.~S.~N. and {Kundu}, R. and {Lares}, M. and {Libralato}, M. and {Lima}, E. and {Maccarone}, T.~J. and {Marchant Cort{\'e}s}, P. and {Martin}, E.~L. and {Masetti}, N. and {Matsunaga}, N. and {Mauro}, F. and {McDonald}, I. and {Mej{\'\i}as}, A. and {Mesa}, V. and {Milla-Castro}, F.~P. and {Minniti}, J.~H. and {Moni Bidin}, C. and {Montenegro}, K. and {Morris}, C. and {Motta}, V. and {Navarete}, F. and {Navarro Molina}, C. and {Nikzat}, F. and {Nilo Castell{\'o}n}, J.~L. and {Obasi}, C. and {Ortigoza-Urdaneta}, M. and {Palma}, T. and {Parisi}, C. and {Pena Ram{\'\i}rez}, K. and {Pereyra}, L. and {Perez}, N. and {Petralia}, I. and {Pichel}, A. and {Pignata}, G. and {Ram{\'\i}rez Alegr{\'\i}a}, S. and {Rojas}, A.~F. and {Rojas}, D. and {Roman-Lopes}, A. and {Rovero}, A.~C. and {Saroon}, S. and {Schmidt}, E.~O. and {Schr{\"o}der}, A.~C. and {Schultheis}, M. and {Sgr{\'o}}, M.~A. and {Solano}, E. and {Soto}, M. and {Stecklum}, B. and {Steeghs}, D. and {Tamura}, M. and {Tissera}, P. and {Valcarce}, A.~A.~R. and {Valotto}, C.~A. and {Vasquez}, S. and {Villalon}, C. and {Villanova}, S. and {Vivanco C{\'a}diz}, F. and {Zelada Bacigalupo}, R. and {Zijlstra}, A. and {Zoccali}, M.},
        title = "{The VISTA Variables in the V{\'\i}a L{\'a}ctea extended (VVVX) ESO public survey: Completion of the observations and legacy}",
      journal = {\aap},
         year = 2024,
        month = sep,
       volume = {689},
          eid = {A148},
        pages = {A148},
          doi = {10.1051/0004-6361/202450584},
archivePrefix = {arXiv},
       eprint = {2406.16646},
 primaryClass = {astro-ph.GA},
       adsurl = {https://ui.adsabs.harvard.edu/abs/2024A&A...689A.148S}
}

@ARTICLE{minniti+2010,
       author = {{Minniti}, D. and {Lucas}, P.~W. and {Emerson}, J.~P. and
         {Saito}, R.~K. and {Hempel}, M. and {Pietrukowicz}, P. and
         {Ahumada}, A.~V. and {Alonso}, M.~V. and {Alonso-Garcia}, J. and
         {Arias}, J.~I. and {Bandyopadhyay}, R.~M. and {Barb{\'a}}, R.~H. and
         {Barbuy}, B. and {Bedin}, L.~R. and {Bica}, E. and {Borissova}, J. and
         {Bronfman}, L. and {Carraro}, G. and {Catelan}, M. and
         {Clari{\'a}}, J.~J. and {Cross}, N. and {de Grijs}, R. and
         {D{\'e}k{\'a}ny}, I. and {Drew}, J.~E. and {Fari{\~n}a}, C. and
         {Feinstein}, C. and {Fern{\'a}ndez Laj{\'u}s}, E. and {Gamen}, R.~C. and
         {Geisler}, D. and {Gieren}, W. and {Goldman}, B. and {Gonzalez}, O.~A. and
         {Gunthardt}, G. and {Gurovich}, S. and {Hambly}, N.~C. and
         {Irwin}, M.~J. and {Ivanov}, V.~D. and {Jord{\'a}n}, A. and
         {Kerins}, E. and {Kinemuchi}, K. and {Kurtev}, R. and
         {L{\'o}pez-Corredoira}, M. and {Maccarone}, T. and {Masetti}, N. and
         {Merlo}, D. and {Messineo}, M. and {Mirabel}, I.~F. and {Monaco}, L. and
         {Morelli}, L. and {Padilla}, N. and {Palma}, T. and {Parisi}, M.~C. and
         {Pignata}, G. and {Rejkuba}, M. and {Roman-Lopes}, A. and
         {Sale}, S.~E. and {Schreiber}, M.~R. and {Schr{\"o}der}, A.~C. and
         {Smith}, M. and {}, L. Sodr{\'e}, Jr. and {Soto}, M. and {Tamura}, M. and
         {Tappert}, C. and {Thompson}, M.~A. and {Toledo}, I. and {Zoccali}, M. and
         {Pietrzynski}, G.},
        title = "{VISTA Variables in the Via Lactea (VVV): The public ESO near-IR variability survey of the Milky Way}",
      journal = {\na},
         year = 2010,
        month = jul,
       volume = {15},
       number = {5},
        pages = {433-443},
          doi = {10.1016/j.newast.2009.12.002},
archivePrefix = {arXiv},
       eprint = {0912.1056},
 primaryClass = {astro-ph.GA},
       adsurl = {https://ui.adsabs.harvard.edu/abs/2010NewA...15..433M}
}

@ARTICLE{Damineli_2016,
       author = {{Damineli}, A. and {Almeida}, L.~A. and {Blum}, R.~D. and {Damineli}, D.~S.~C. and {Navarete}, F. and {Rubinho}, M.~S. and {Teodoro}, M.},
        title = "{Extinction law in the range 0.4-4.8 {\ensuremath{\mu}}m and the 8620 {\r{A}} DIB towards the stellar cluster Westerlund 1}",
      journal = {\mnras},
         year = 2016,
        month = dec,
       volume = {463},
       number = {3},
        pages = {2653-2666},
          doi = {10.1093/mnras/stw2122},
archivePrefix = {arXiv},
       eprint = {1607.04639},
 primaryClass = {astro-ph.SR},
       adsurl = {https://ui.adsabs.harvard.edu/abs/2016MNRAS.463.2653D}
}

@ARTICLE{McInnes2017,
       author = {{McInnes}, Leland and {Healy}, John and {Astels}, Steve},
        title = "{hdbscan: Hierarchical density based clustering}",
      journal = {The Journal of Open Source Software},
         year = 2017,
        month = mar,
       volume = {2},
       number = {11},
          eid = {205},
        pages = {205},
          doi = {10.21105/joss.00205},
       adsurl = {https://ui.adsabs.harvard.edu/abs/2017JOSS....2..205M}
}

@InProceedings{Campello2013,
    author="Campello, Ricardo J. G. B.
    and Moulavi, Davoud
    and Sander, Joerg",
    editor="Pei, Jian
    and Tseng, Vincent S.
    and Cao, Longbing
    and Motoda, Hiroshi
    and Xu, Guandong",
    title="Density-Based Clustering Based on Hierarchical Density Estimates",
    booktitle="Advances in Knowledge Discovery and Data Mining",
    year="2013",
    publisher="Springer Berlin Heidelberg",
    address="Berlin, Heidelberg",
    pages="160--172",
    isbn="978-3-642-37456-2"
}

@ARTICLE{Smith_2025,
       author = {{Smith}, Leigh C. and {Lucas}, Philip W. and {Koposov}, Sergey E. and {Gonzalez-Fernandez}, Carlos and {Alonso-Garc{\'\i}a}, Javier and {Minniti}, Dante and {Sanders}, Jason L. and {Bedin}, Luigi R. and {Belokurov}, Vasily and {Evans}, N. Wyn and {Hempel}, Maren and {Ivanov}, Valentin D. and {Kurtev}, Radostin G. and {Saito}, Roberto K.},
        title = "{VIRAC2: NIR astrometry and time series photometry for 500M+ stars from the VVV and VVVX surveys}",
      journal = {\mnras},
         year = 2025,
        month = feb,
       volume = {536},
       number = {4},
        pages = {3707-3738},
          doi = {10.1093/mnras/stae2797},
archivePrefix = {arXiv},
       eprint = {2501.06295},
 primaryClass = {astro-ph.GA},
       adsurl = {https://ui.adsabs.harvard.edu/abs/2025MNRAS.536.3707S}
}

\begin{appendix} 
\onecolumn


\section{Dipper variables}
\label{dippers_appendix}
Variability parameters of the members classifies as dippers and confirmed as this class by visual inspection. Table \ref{table:dippers} lists all $44$ dipper stars and contains their VIRAC2 IDs (sourceid column in its main source catalog) their equatorial coordinates, the obtained Lomb-Scargle period, the amplitude of variation $\Delta K_{s}$, the $Q$ and $M$ metrics, the mean $K_{\rm{s}}$ magnitude and error and the number of epochs in each of their light curves $N$.\\ 

Figs.~\ref{dippers1}, \ref{dippers2}, \ref{dippers3} and \ref{dippers4} show the phase-folded light curves of stars with dips, confirmed by visual inspection. Their VIRAC2 IDs and periods are shown as a title on each plot. Period values should be further confirmed.

\begin{table*}[ht!]
\caption{Variability analysis of dipper light curves}             
\label{table:dippers}      
\centering          
\begin{tabular}{l c c c c c c c c c }     
\hline\hline       
sourceid & $\alpha$ & $\delta$ & $P$ & $\Delta K_{s}$ & $Q$ & $M$ & $\overline{K_{s}}$ & $\overline{eK_{s}}$ & $N$ \\
- & [degrees] & [degrees] & [d] & [mag] &  &  & [mag] & [mag] &  \\ 
\hline                    
14999470021717 & 251.8427 & -45.8311 & 1.244 & 0.133 & 0.505 & 0.819 & 11.087 & 0.017 & 103 \\
14999470001971 & 251.7899 & -45.7967 & 1.897 & 0.206 & 0.538 & 0.989 & 10.828 & 0.017 & 87 \\
14995696009595 & 251.7548 & -45.7947 & 0.634 & 0.198 & 0.747 & 0.828 & 14.485 & 0.026 & 128 \\
14999470013838 & 251.8045 & -45.8239 & 7.702 & 0.241 & 0.471 & 0.694 & 14.875 & 0.040 & 127 \\
14999469019626 & 251.7493 & -45.8332 & 0.413 & 0.143 & 0.494 & 0.709 & 12.285 & 0.020 & 75 \\
14999469013165 & 251.7332 & -45.8695 & 1.217 & 0.129 & 0.438 & 0.559 & 12.165 & 0.021 & 119 \\
14999469005518 & 251.7657 & -45.8325 & 0.411 & 0.168 & 0.382 & 0.614 & 10.771 & 0.026 & 58 \\
15003238011361 & 251.6819 & -45.8757 & 0.683 & 0.153 & 0.756 & 1.018 & 11.174 & 0.021 & 97 \\
14999469001399 & 251.7260 & -45.8506 & 1.843 & 0.215 & 0.742 & 0.910 & 14.768 & 0.038 & 124 \\
14999469009992 & 251.7455 & -45.8556 & 1.784 & 0.133 & 0.270 & 0.689 & 13.810 & 0.025 & 125 \\
15003239021845 & 251.8025 & -45.8777 & 0.448 & 0.326 & 0.456 & 1.189 & 13.986 & 0.024 & 128 \\
14999469015611 & 251.7282 & -45.8432 & 0.630 & 0.263 & 0.760 & 0.681 & 14.953 & 0.044 & 127 \\
15007003001087 & 251.7180 & -45.9377 & 1.190 & 0.325 & 0.603 & 1.102 & 12.081 & 0.019 & 123 \\
15007003004473 & 251.6748 & -45.9383 & 0.481 & 0.288 & 0.597 & 0.562 & 14.855 & 0.039 & 175 \\
15003238005721 & 251.6946 & -45.8554 & 0.413 & 0.226 & 0.501 & 0.562 & 14.949 & 0.036 & 126 \\
15003238001552 & 251.7367 & -45.8898 & 0.671 & 0.776 & 0.668 & 1.044 & 15.153 & 0.056 & 176 \\
14999469014074 & 251.7268 & -45.8559 & 7.726 & 0.186 & 0.727 & 0.738 & 14.277 & 0.026 & 127 \\
15003239023448 & 251.7794 & -45.8959 & 0.887 & 0.218 & 0.590 & 0.622 & 14.656 & 0.031 & 127 \\
14999469016226 & 251.7352 & -45.8686 & 0.481 & 0.213 & 0.348 & 0.729 & 15.017 & 0.047 & 124 \\
14999469018973 & 251.7228 & -45.8608 & 1.271 & 0.473 & 0.707 & 0.716 & 15.201 & 0.056 & 122 \\
14999469012388 & 251.7117 & -45.8145 & 1.171 & 0.425 & 0.756 & 0.970 & 15.335 & 0.052 & 127 \\
15007004002919 & 251.7554 & -45.9180 & 3.791 & 0.152 & 0.648 & 0.911 & 11.199 & 0.020 & 90 \\
14999469010623 & 251.7448 & -45.8267 & 0.498 & 0.323 & 0.706 & 0.706 & 14.926 & 0.041 & 126 \\
15003238003010 & 251.7033 & -45.8978 & 1.121 & 0.129 & 0.259 & 0.718 & 11.426 & 0.020 & 101 \\
15007004006159 & 251.7793 & -45.9357 & 0.652 & 0.542 & 0.683 & 0.799 & 14.635 & 0.030 & 128 \\
14999469007702 & 251.7527 & -45.8519 & 2.562 & 0.147 & 0.619 & 1.054 & 12.630 & 0.024 & 72 \\
15003238007188 & 251.6596 & -45.8837 & 1.303 & 0.121 & 0.524 & 1.087 & 11.199 & 0.021 & 94 \\
15003238012223 & 251.6870 & -45.8529 & 1.062 & 0.137 & 0.636 & 0.880 & 11.377 & 0.020 & 104 \\
15003239014379 & 251.7937 & -45.8623 & 1.452 & 0.222 & 0.621 & 0.565 & 14.403 & 0.034 & 186 \\
15007004006805 & 251.7679 & -45.8962 & 1.843 & 0.127 & 0.448 & 0.662 & 14.098 & 0.029 & 124 \\
15003238009456 & 251.7214 & -45.9084 & 0.554 & 0.174 & 0.662 & 0.678 & 11.552 & 0.021 & 160 \\
15007004006894 & 251.7573 & -45.9093 & 3.055 & 0.292 & 0.688 & 0.624 & 15.326 & 0.048 & 126 \\
14999469014013 & 251.7621 & -45.8540 & 0.425 & 0.165 & 0.521 & 0.674 & 11.238 & 0.022 & 49 \\
15003238004148 & 251.7183 & -45.8702 & 2.187 & 0.309 & 0.557 & 0.704 & 14.993 & 0.039 & 125 \\
14999469009227 & 251.6941 & -45.8421 & 1.661 & 0.569 & 0.764 & 0.564 & 14.036 & 0.039 & 88 \\
15003237008756 & 251.6224 & -45.8951 & 0.548 & 0.290 & 0.680 & 0.664 & 15.047 & 0.042 & 125 \\
15003237001059 & 251.6077 & -45.8596 & 0.820 & 0.434 & 0.471 & 0.709 & 11.177 & 0.021 & 140 \\
14999468004612 & 251.6610 & -45.8401 & 1.111 & 0.146 & 0.571 & 0.585 & 10.957 & 0.022 & 78 \\
14995695014488 & 251.6186 & -45.7617 & 1.701 & 0.210 & 0.720 & 0.599 & 14.287 & 0.027 & 125 \\
15003237008281 & 251.6064 & -45.9002 & 0.729 & 0.340 & 0.762 & 0.767 & 15.365 & 0.047 & 123 \\
14999468002615 & 251.6373 & -45.8324 & 0.514 & 0.184 & 0.679 & 0.761 & 11.301 & 0.022 & 143 \\
14999468007854 & 251.6340 & -45.8065 & 0.517 & 0.131 & 0.596 & 0.809 & 10.810 & 0.023 & 71 \\
14991918004195 & 251.6924 & -45.7331 & 2.215 & 0.171 & 0.645 & 0.566 & 14.554 & 0.028 & 126 \\
14988137003990 & 251.6921 & -45.7117 & 0.885 & 0.188 & 0.534 & 0.709 & 14.668 & 0.030 & 128 \\
\hline                  
\end{tabular}
\end{table*}

\begin{figure*}[ht!]
\centering
\includegraphics[width=\textwidth]{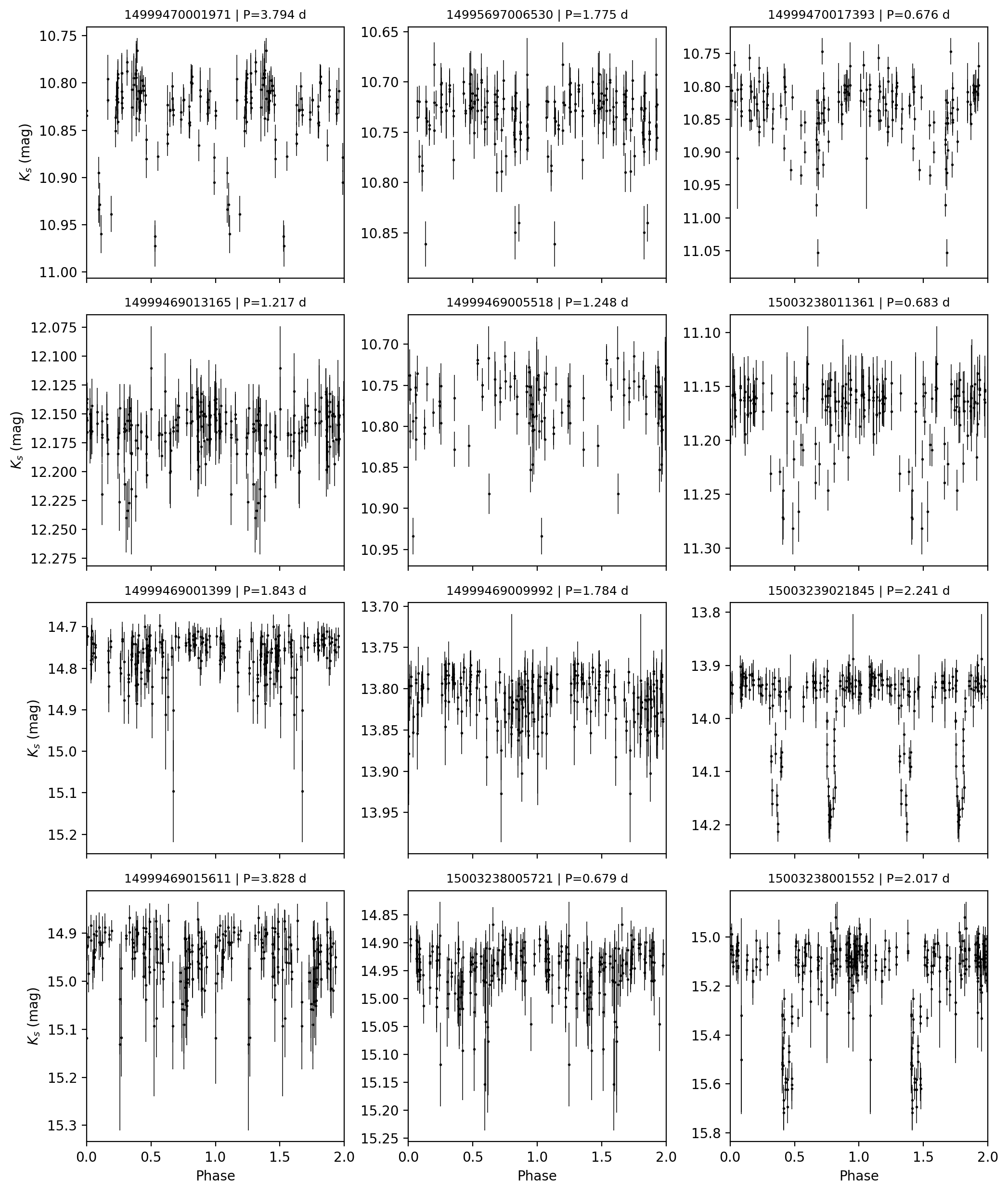}
\caption{Phase folded light curves of dipper stars. VIRAC2 IDs and the period $P$ of each are shown.}
\label{dippers1}
\end{figure*}

\begin{figure*}
\centering
\includegraphics[width=\textwidth]{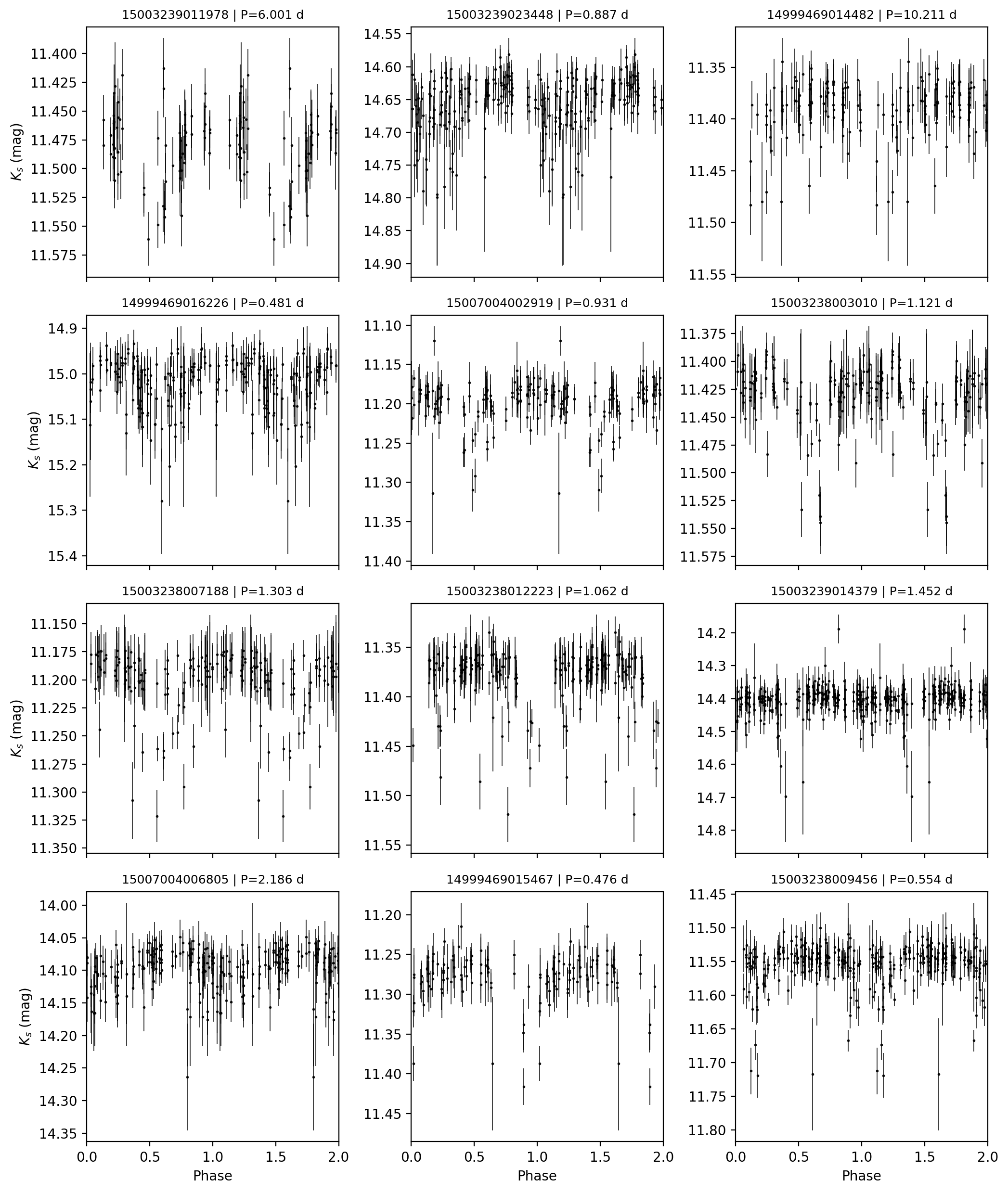}
\caption{Continued.}
\label{dippers2}
\end{figure*}

\begin{figure*}
\centering
\includegraphics[width=\textwidth]{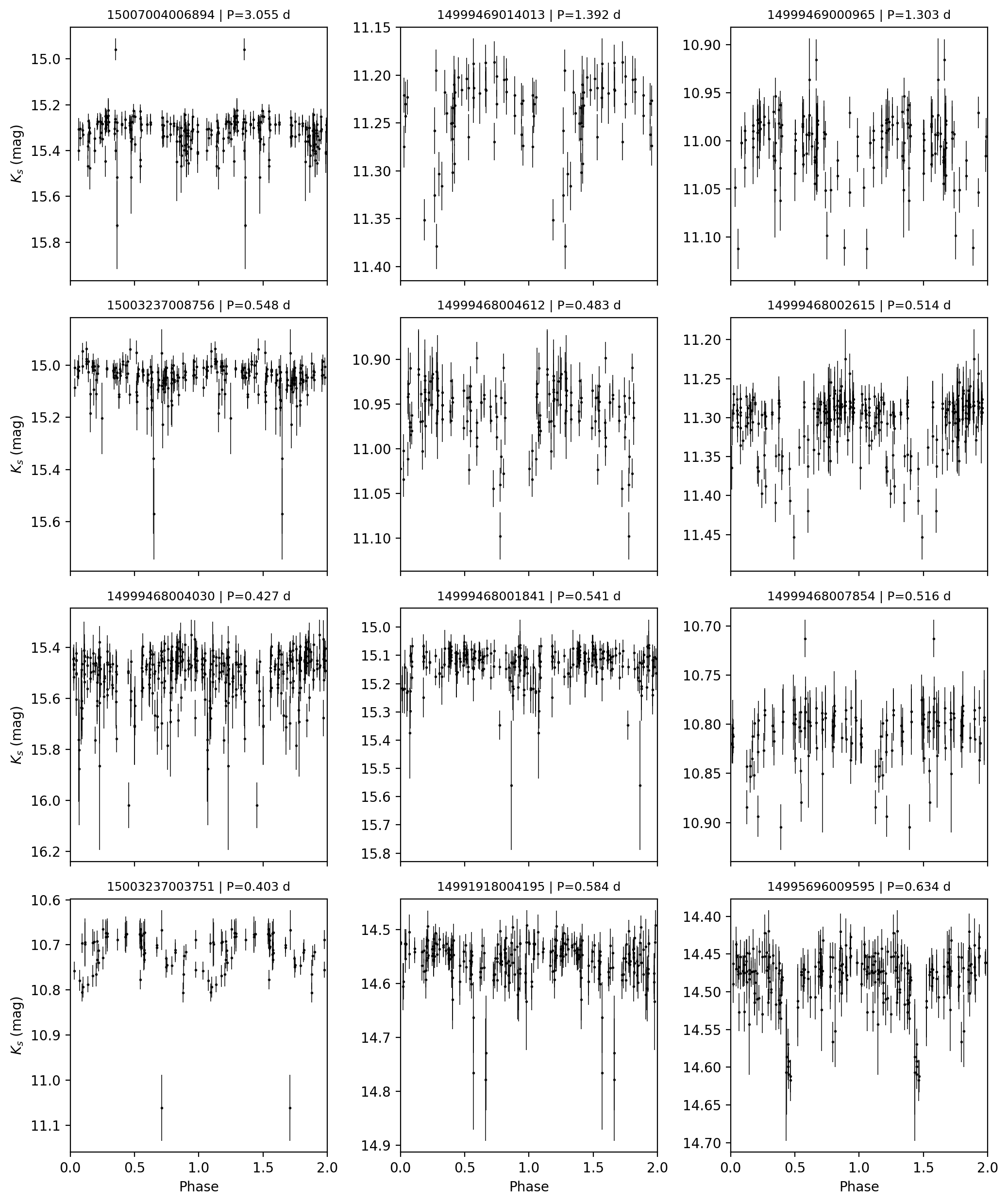}
\caption{Continued.}
\label{dippers3}
\end{figure*}

\begin{figure*}
\centering
\includegraphics[width=\textwidth]{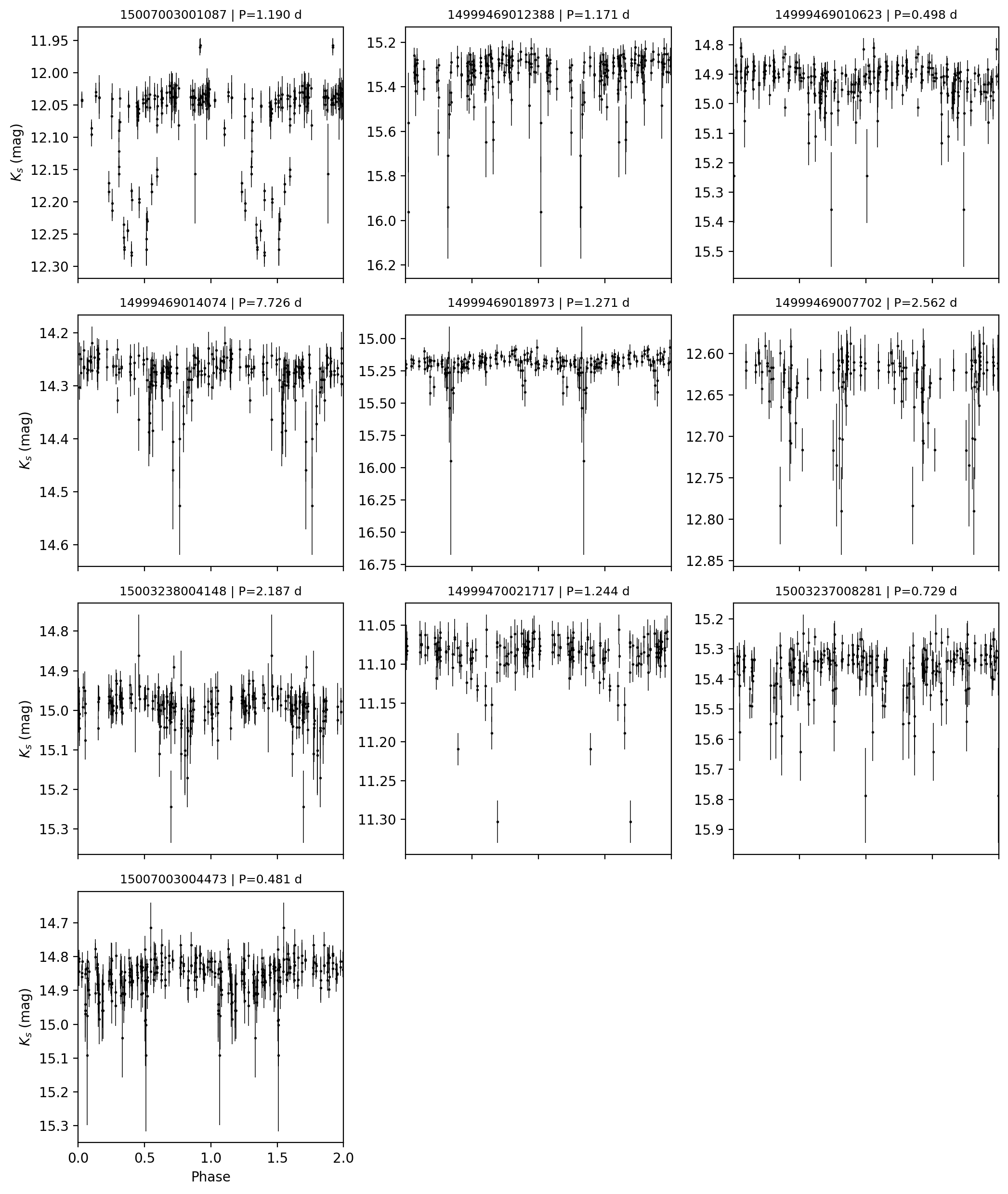}
\caption{Continued.}
\label{dippers4}
\end{figure*}

\end{appendix}

\end{document}